\documentclass[aps,prl,twocolumn,nofootinbib,groupedaddress,superscriptaddress,secnumarabic,
,longbibliography]{revtex4-2}
\AtBeginDocument{
\heavyrulewidth=.08em
\lightrulewidth=.05em
\cmidrulewidth=.03em
\belowrulesep=.65ex
\belowbottomsep=0pt
\aboverulesep=.4ex
\abovetopsep=0pt
\cmidrulesep=\doublerulesep
\cmidrulekern=.5em
\defaultaddspace=.5em
}

\usepackage{float}
\usepackage{comment}
\usepackage{booktabs}
\usepackage[percent]{overpic}
\usepackage{microtype}
\usepackage{array}
\usepackage{physics}
\usepackage{graphicx}
\usepackage{bbold}
\usepackage[usenames,dvipsnames]{xcolor}
\usepackage{tikz}
\usepackage{epsfig}

\definecolor{ForestGreen}{HTML}{668000}
\definecolor{red1}{HTML}{FF4136}
\definecolor{green1}{HTML}{00802b}	
\definecolor{tum}{HTML}{0065bd}

\usepackage{amsmath,amssymb}
\usepackage{multirow}
\usepackage{bm}
\usepackage{mathtools}
\usepackage{dsfont}
\usepackage{amsfonts}
\usepackage[normalem]{ulem}
\usepackage{url}
\usepackage{wasysym}

\usepackage[colorlinks=true,linkcolor=tum,citecolor=tum,urlcolor=tum, hypertexnames=false]{hyperref}

\def\ba#1\ea{\begin{align}#1\end{align}}
\def\bg#1\eg{\begin{gather}#1\end{gather}}
\def\bpm{\begin{pmatrix}}
\def\epm{\end{pmatrix}}

\definecolor{myBlue}{HTML}{0065BD}   
\newcommand{\tum}[1]{\textcolor{myBlue}{#1}}

\newcommand{\ourtitle}{Transient localization from fractionalization: \\ vanishingly small heat conductivity in gapless quantum magnets}

\allowdisplaybreaks

\begin{document}
\title{\textbf{\ourtitle}}

\author{Shi Feng}
\email{shi.feng@tum.de}
\affiliation{Technical University of Munich, TUM School of Natural Sciences, Physics Department, 85748 Garching, Germany}
\affiliation{Munich Center for Quantum Science and Technology (MCQST), Schellingstr. 4, 80799 M{\"u}nchen, Germany}

\author{Penghao Zhu}
\affiliation{Department of Physics, The Ohio State University, Columbus, Ohio 43210, United States}

\author{Johannes Knolle}
\affiliation{Technical University of Munich, TUM School of Natural Sciences, Physics Department, 85748 Garching, Germany}
\affiliation{Munich Center for Quantum Science and Technology (MCQST), Schellingstr. 4, 80799 M{\"u}nchen, Germany}
\affiliation{Blackett Laboratory, Imperial College London, London SW7 2AZ, United Kingdom}

\author{Michael Knap}
\affiliation{Technical University of Munich, TUM School of Natural Sciences, Physics Department, 85748 Garching, Germany}
\affiliation{Munich Center for Quantum Science and Technology (MCQST), Schellingstr. 4, 80799 M{\"u}nchen, Germany}

\begin{abstract}
Several candidate materials for gapless quantum spin liquids exhibit a vanishing thermal conductivity, which is at odds with theoretical predictions. Here, we show that a suppressed response can arise due to transient localization from fractionalization, even in the absence of extrinsic defects or disorder. Concretely, we consider a Kitaev ladder model in a uniform magnetic field, whose spin degrees of freedom fractionalize into visons and spinons. For moderate magnetic fields,   visons are heavy and act as quasi-static disorder that induce transient localization of light spinons even in the translation-invariant model and at zero temperature, which strongly suppresses the residual conductivity at finite but low frequencies. At ultralow frequencies the conductivity is restored; however, such scales can be extremely hard to reach in experiments. Our results identify transient localization as a signature of fractionalization and provide a framework for interpreting anomalous transport in gapless spin liquid candidates.
\end{abstract}

\maketitle
\let\oldaddcontentsline\addcontentsline
\renewcommand{\addcontentsline}[3]{}

{\it \tum {Introduction.--} }The description of phases of matter and their response theory is typically formulated in terms of long-wavelength effective theories and provides a unifying framework for understanding emergent phenomena. For systems with fractionalized excitations, these descriptions are typically obtained from parton constructions and associated gauge theories, where the emergent gauge structure is assumed to be translationally invariant~\cite{wen_book}. Within this framework, transport and thermodynamic properties are governed by the long-wavelength dynamics of emergent quasi-particles and gauge fluctuations.

Despite the success of this paradigm in capturing key qualitative aspects of many fractionalized phases \cite{savary2016quantum,broholm2020quantum,Knolle_ARCMP_2019}, challenges arise when theoretical predictions are compared to several experiments. For example, gapless quantum spin liquids (QSL) with emergent Fermi surfaces are predicted to have thermodynamic signatures identical to a metal: a linear-in-temperature specific heat and thermal conductivity~\cite{balents2010spin}. Several frustrated Mott insulators exhibit such linear specific heat down to the lowest temperatures \cite{Pengcheng21,Yamashita2011,Zhang24,matsuda2025}.  Yet, various compounds, such as candidates of field-stabilized Kitaev QSL: $\alpha$-RuCl$_3$ \cite{nokappaKitaev18} and Na$_2$Co$_2$TeO$_6$ \cite{Hong2024}; geometrically frustrated spin liquids on the triangular lattice like YbMgGaO$_4$ \cite{Xu16} and NaYbSe$_2$ \cite{Zhang24}; and the organic spin liquid candidates \cite{Taillefer19,Li19}, show a concomitant \emph{vanishing} thermal conductivity at low temperatures, at variance with the theoretical expectation. Therefore, refinements of the long-wavelength effective theories might be required to understand the experimental implications. The coexistence of linear specific heat and seemingly vanishing residual conductivity challenges the presence of gapless quantum spin liquids, and is frequently ascribed to extrinsic defects and disorder. 

Here, we show that this paradox can arise intrinsically from fractionalization in the absence of extrinsic defects or disorder, and that the very mechanism which produces a finite spinon density at zero energy also suppresses residual conductivity, rendering it a hallmark of fractionalization.
Concretely, we demonstrate the idea in a translation-invariant Kitaev model on a ladder geometry in which a moderate uniform magnetic field drives a gapped spinon band insulator into a gapless fractionalized phase with light spinons and heavy visons that mutually interact~\cite{kitaev2006anyons}. The heavy visons imprint a quasi-static coherent randomness on the spinons, thereby localizing them transiently \emph{without quenched or extrinsic disorder}~\cite{Schiulaz2014,DeRoeck2014,Papic2015,Moore16,DarkwahOppong2020}. As a consequence, the heat conductivity will be strongly suppressed as frequency is lowered. Asymptotically at ultralow energy scales, the system will become conducting, requiring, however, experimental probes under extreme conditions and the absence of any extrinsic perturbations. The main mechanism for the observed suppressed heat conductivity is therefore the transient localization (TL) of spinons induced by the heavy visons.

\begin{figure}[t]
    \centering
    \includegraphics[width=\linewidth]{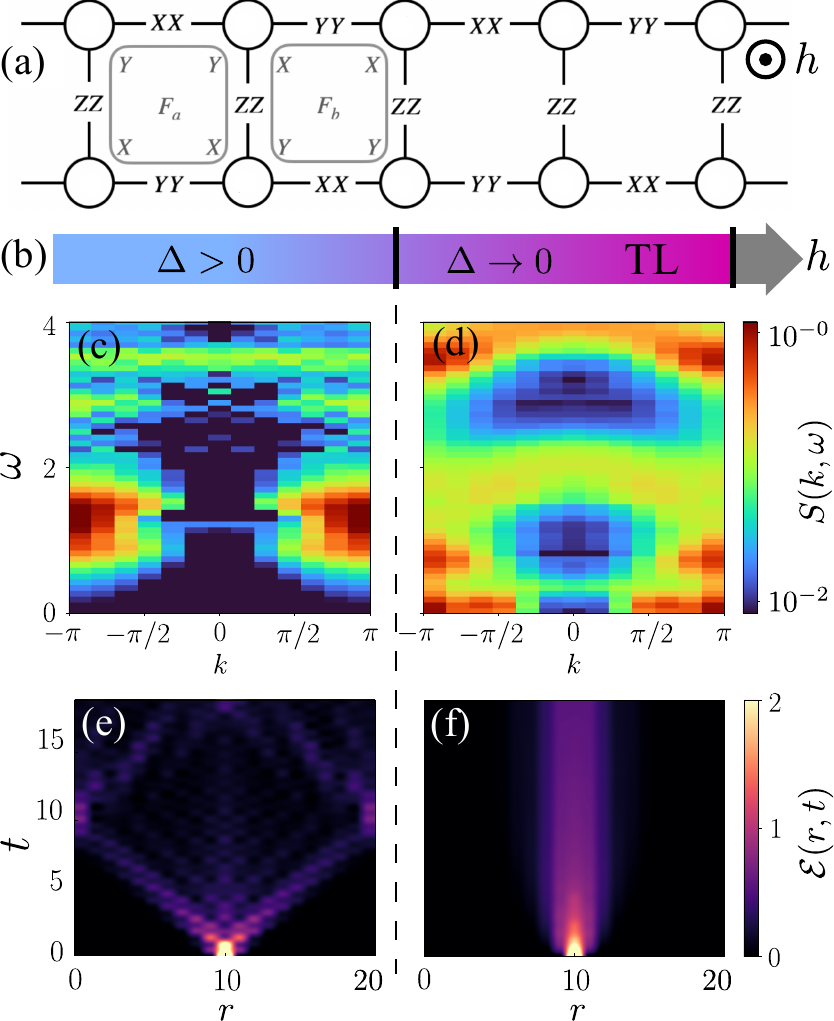}
    \caption{Model and numerical results for field-induced gapless spectrum and transient localization (TL) from fractionalization.  (a) The Kitaev ladder model with local conservation laws $F_a$ and $F_b$. (b) Schematic phase diagram of Hamiltonian~\eqref{eq:Ham}: increasing the magnetic field $h$ drives a transition from the gapped spin liquid (blue) at $h\approx 0.4$, to a gapless regime with TL (purple), and then to the partially polarized phase (gray) at $h \approx 0.9$. The blue-to-purple gradient in (b) denotes the increasing vison filling factor $\nu$.  (c,d) Spin spectral function at $h=0.04$ (c) and $h=0.80$ (d). (e,f) Energy-density spreading after a local quench applied to the ground state for $h=0.04$ (e) and $h=0.80$ (f). 
    }
    \label{fig:lattice}
\end{figure}

{\it \tum {The Kitaev ladder.-- }} 
We start from a Kitaev ladder model of local spin-$\frac{1}{2}$ moments in the presence of a field:
\begin{align}
    H &= H_{\rm ladder} - h \sum_{j,l} (X_j^l + Y_j^l + Z_j^l) \label{eq:Ham}\\
    H_{\rm ladder} &= \sum_{j + l \in \rm even} \left(X_{j}^l X_{j+1}^l + Y_{j-1}^l Y_{j}^l + Z_{j}^l Z_{j}^{l+1} \right) \label{eq:Ham0}.
\end{align}
Here, the subscript $j$ is the horizontal index along the ladder, and the superscript $l$ a binary index labeling the lower ($l=0$) or upper ($l=1$) leg, see Fig.~\ref{fig:lattice}(a). 
Similarly to the Kitaev's honeycomb model, $H_{\rm ladder}$ has extensive local symmetries given by the square faces $F_a = X^0_0 Y^1_0 X^0_1 Y^1_1$ and $F_b = Y^0_1 X^1_1 Y^0_2 X^1_2$, with $F_{a(b)} = 1$ in the ground state \cite{kitaev2006anyons, Xiang07}.  
For $h\rightarrow 0$, elementary excitations of $H_{\rm ladder}$ are free Majorana fermions (referred to as spinons) and static $Z_2$ fluxes (visons) given by $F_{a(b)} = -1$. In contrast to the honeycomb model, the Majorana band in the ground state sector of $H_{\rm ladder}$ is highly gapped, see supplemental materials \cite{supp}. A non-zero magnetic field breaks the exact local conservation laws. For a small-to-moderate $h$, the ground-state expectation of $F_{a(b)}$ decreases from $1.0$ to $\sim 0$ \cite{Baskaran2023,Feng2023,wang2024}, indicating an increasing density of flipped fluxes $\nu \sim -\frac{1}{2}(\expval{F_{a(b)}} - 1)$ at low energy to near half filling, as illustrated by the blue-to-purple gradient in the cartoon phase diagram Fig.~\ref{fig:lattice}(b). The finite flux density $\nu$ under moderate $h$ allows us to put forward a simple approximate description of the ground state as a superposition of quasi-static vison configurations. Since each typical configuration is disordered the fermions are localized~\cite{Cirac05} up to a cut-off timescale set by the weak tunneling between different vison configurations~\cite{Schiulaz2014, DeRoeck2014, Papic2015, Moore16, DarkwahOppong2020}. 
We next present numerical evidence using tensor network approaches to support this effective theory. We demonstrate the presence of transient localization (TL) by suppressed energy transport, and the simultaneous presence of fractionalized gapless excitations.

{\it \tum {Coexistence of vanishing gap and slow dynamics.--} }
Beyond the weak-field limit, the interplay between Majorana fermions and proliferating visons becomes important, leading to a transition into a gapless region under moderate field. 
This is reflected in the dynamical spin structure factor shown in Fig.~\ref{fig:lattice}(c,d) obtained by the density matrix renormalization group (DMRG) \cite{White99,Alvarez16}: while for small but finite field $h=0.04$ there is an observable gap $\Delta$ due to gapped Majorana band [Fig.~\ref{fig:lattice}(c)], at moderate field $h=0.8$ there are concentrated gapless modes around $k=\pm \pi$ points [Fig.~\ref{fig:lattice}(d)]. 
Qualitatively, this is similar to the emergence of the finite zero-energy density of states of fermions due to thermal visons \cite{Nasu15}; however, in this case it is rather the quantum vison fluctuation that is responsible for the drastic change in the zero-temperature structure factor \cite{wang2024,penghao24}, which we will elaborate later. 

Now, we show the coexisting slow dynamics in this gapless regime. 
In our model, the energy is the only conserved global charge of Eq.~\eqref{eq:Ham} at finite $h$. We therefore demonstrate the coexisting slow dynamics first by measuring the spreading of energy-density fluctuation $\mathcal{E}_r = h_r - \expval{h_r}$, where $h_r$ is the local Hamiltonian of the $r$-th unit cell. 
Since we are interested in the low-energy dynamics, the initial state $\ket{0'}$ is prepared by perturbing the ground state $\ket{0}$ obtained from DMRG, locally by $P_{r_0} = Z_{r_0}^0 Z_{r_0}^1$, where $r_0$ is at the center of the ladder.
Results are shown in Fig.~\ref{fig:lattice}(e,f), obtained by matrix product state (MPS) time evolution from $\ket{0'}$ (with $t_{\max} \sim 20$ and bond dimensions up to $1000$)~\cite{tenpy2024}.
For $h\rightarrow 0$, $P_{r_0}$ is $Z_2$ gauge-invariant, perturbing only Majorana fermions. Thus there is a fast spread of energy density for $h = 0.04$ [Fig.~\ref{fig:lattice}(e)]. The intensity at the center of the ladder decays on a very short timescale set by $1/\rm bandwidth$; and the correlation spreads as a lightcone (see supplemental materials \cite{supp}). By contrast, for the moderate field $h=0.8$, there is no obvious energy spreading beyond the early time dynamics [Fig.~\ref{fig:lattice}(f)]. The energy density is localized within a length scale of $\xi \sim 8$ unit cells \cite{supp}.  Similar slow dynamics under moderate field can also be seen in the correlation spreading \cite{supp}.  This indicates that even a translation-invariant gapless fractionalized phase can have ultraslow transport, thus, exhibiting TL \cite{Schiulaz2014, DeRoeck2014, Papic2015, Moore16, DarkwahOppong2020} in the low energy sectors. 

\begin{figure}[t]
    \centering
    \includegraphics[width=\linewidth]{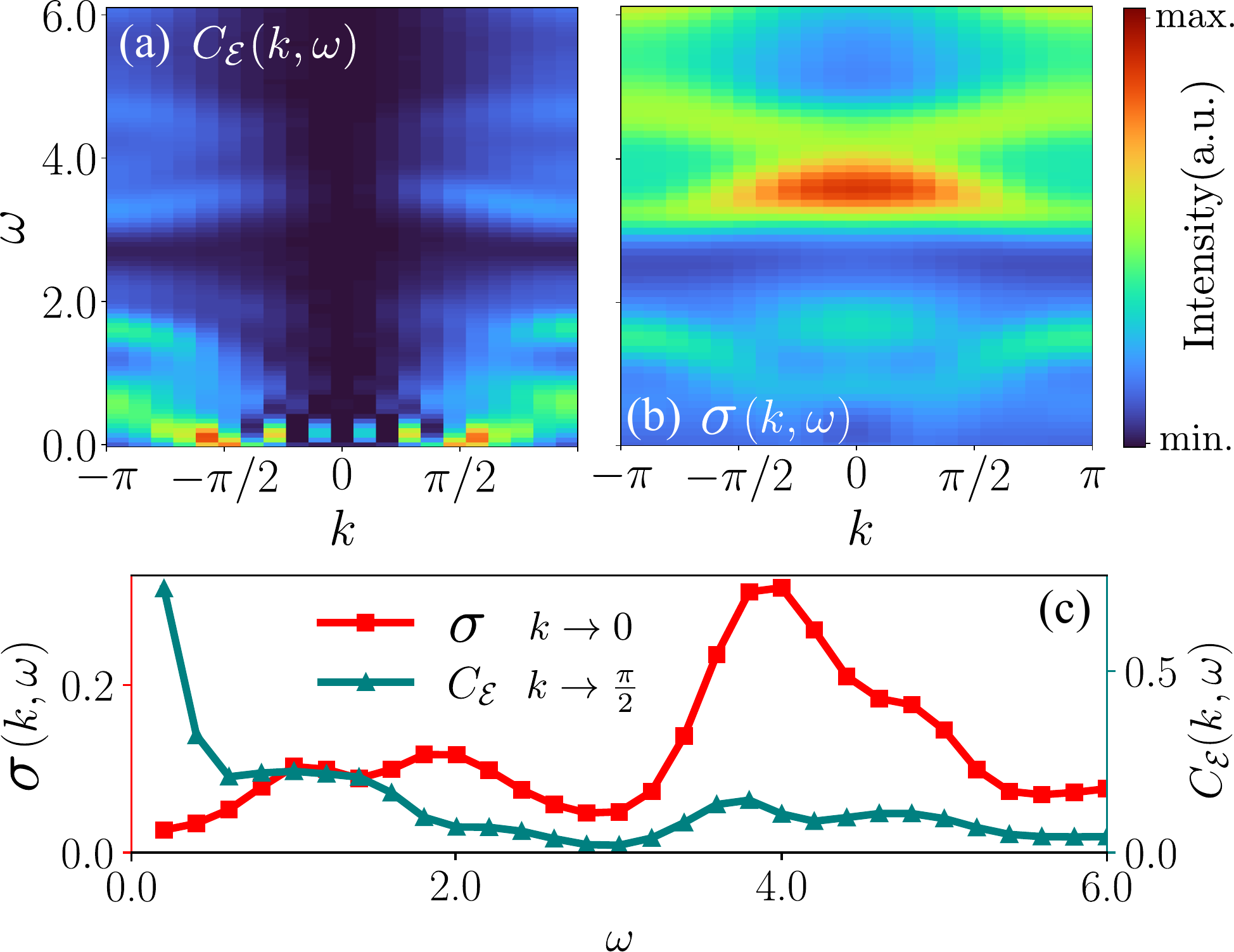}
    \caption{Coexistence of a gapless energy‐fluctuation spectrum and a virtually gapped energy‐current spectrum in the TL regime. (a) Dynamical energy correlation function $C_\mathcal{E}(k, \omega)$ and (b) energy conductivity $\sigma(k, \omega)$ at $h=0.8$. Intensities are plotted on linear scale. (c) Right axis: Cut of $C_\mathcal{E}(k, \omega)$ at $k\rightarrow \frac{\pi}{2}$ (teal); Left axis: cut of $\sigma(k, \omega)$ at $k\to 0$ (red).}
    \label{fig:blurb}
\end{figure}

{\it \tum {Suppressed energy conduction.--} }
We provide further evidence for the TL by studying the energy current response, which is directly measurable in experiments via thermal transport. 
Using the definition of a local energy-density operator and the discrete continuity equation $\frac{d}{dt} \mathcal{E}(r,t) + \left[J_{r, r+1}(t) - J_{r-1, r}(t) \right]= 0$, where $J$ is the energy current operator, we obtain, see supplement~\cite{supp}:
\begin{equation} \label{eq:jjnn}
    C_J(k, \omega) = \frac{\omega^2 \,C_\mathcal{E}(k, \omega)}{2 - 2 \cos(k)},\;\;\; \sigma(k, \omega) \sim \frac{C_J(k, \omega)}{\omega},
\end{equation}
which relates the two correlation functions $C_\mathcal{E} = \expval{\mathcal{E}_k(\omega) \mathcal{E}_{-k}(\omega)}$ and $C_J = \expval{J_k(\omega) J_{-k}(\omega)}$, reflecting respectively the spectral density of energy and current; and $\sigma(k, \omega)$ the heat conductivity \cite{Luttinger64}.
Consistency is verified by confirming that the independently computed $C_J(k,\omega)$ coincides with $C_{\mathcal E}(k,\omega)$ multiplied by $\omega^2/(2-2\cos k)$, see supplemental materials \cite{supp}.
Both $C_\mathcal{E}(k, \omega)$ and $\sigma(k, \omega)$ at a moderate field $h=0.8$ are shown in Fig.~\ref{fig:blurb}. 
$C_\mathcal{E}$ in Fig.~\ref{fig:blurb}(a) has significant gapless spectral weight especially about $k_0 = \pm \frac{\pi}{2}$. 
This again confirms the presence of gapless excitations, consistent with the spin spectral function in Fig.~\ref{fig:lattice}(d). 
However, in the numerically accessible energy scales, $\sigma(k, \omega)$, as well as $C_J(k, \omega)$, are suppressed at all momenta for small frequencies, see Fig.~\ref{fig:blurb}(b,c) and supplemental material \cite{supp}. Such a suppression in the DC-limit of the conductivity in a gapless system is a hallmark of TL \cite{Giorgio06,Heller24,Hadi24}, in contrast to the infrared Lorentzian peak in diffusive systems.

\begin{figure*}[t]
    \centering
    \includegraphics[width=\linewidth]{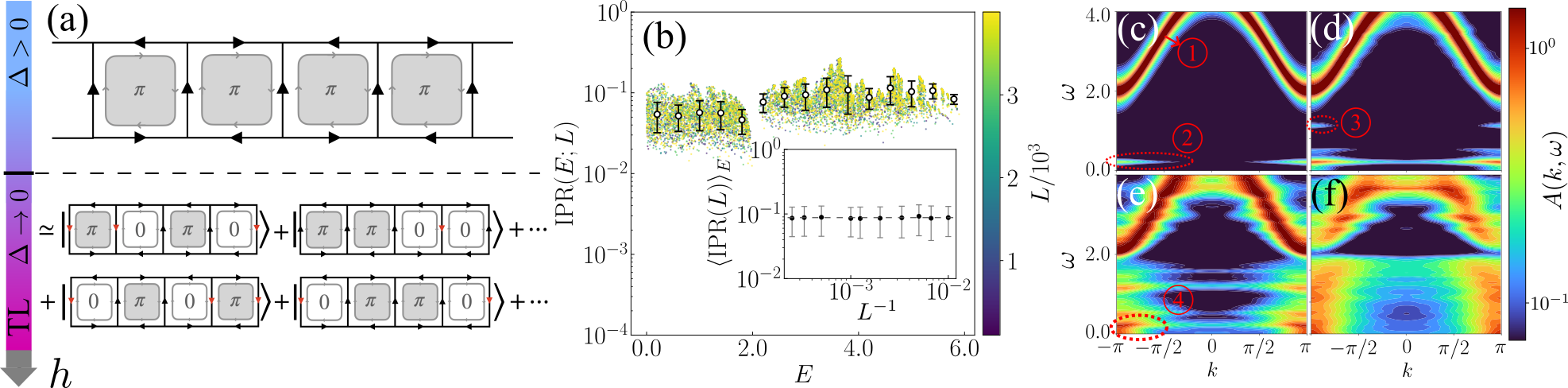}
    \caption{Illustration and results of the effective theory with quasi-static approximation. (a) Left: schematic phase diagram; top: ground-state sector with uniform $\pi$ flux for $h\to 0$, black arrows denote the $Z_2$ gauge; bottom: field-induced superposition of quasi-static fluxes, with red arrows denoting the flipped gauge links with respect to the $\pi$-flux ladder at $h\to 0$.  (b) Energy-resolved IPR of the effective tight-binding model conditioned on a flux sector with $\nu = 0.5$; the inset shows the size-independent IPR averaged over the whole bandwidth. (c-f)
    Majorana spectral function for increasing propability $p=p(\nu)$ of flipped flux pairs: (c) $p=0.04$, (d) $p=0.1$, (e) $p=0.3$, and (f) $p=0.5$, resolved in average momentum. The circled labels denote contributions of: \raisebox{.5pt}{\textcircled{\raisebox{-.9pt} {1}}} bulk spinon band, \raisebox{.5pt}{\textcircled{\raisebox{-.9pt} {2}}} in-gap Majorana resonances, \raisebox{.5pt}{\textcircled{\raisebox{-.9pt} {3}}} higher-order Majorana resonances, \raisebox{.5pt}{\textcircled{\raisebox{-.9pt} {4}}} emergent gapless spectral weight.  }
    \label{fig:gapless}
\end{figure*}

{\it \tum {Transient localization by quasi-static visons.-- }}Having established the phenomenology numerically, we next discuss the underlying mechanism.  
Starting from zero field,
$H_{\rm ladder}$ can be fermionized without introducing redundancies: $H_{\rm ladder} = i \sum_{j+l \in \rm even} \left( c_{j+1}^l c_j^l + c_{j-1}^l c_j^l + D_{jl} c_j^{l+1} c_{j}^l\right)$ with $c$ the Majorana operator, and $D_{jk} = \pm 1$ a static local $Z_2$ field defined on vertical bonds. The ground-state gauge configuration is presented in the top row of Fig.~\ref{fig:gapless}(a), with $D_{jk}$ on vertical bonds defined in consistency with a uniform $\pi$-flux configuration, see supplemental material \cite{supp}. 
The ground state is thus of the form $\ket{0} = \ket{M_D} \otimes \ket{D}$ with $\ket{M_D}$ the fermion sector conditioned on the static fluxes given by $\ket{D}$. 
Upon turning on $h$, the ground state is no longer the exact $\pi$-flux ladder, rather, it consists of a superposition of multiple gauge-superselection sectors hybridized with the fermions. The mixed dynamics of visons and fermions is difficult to analyze analytically. However, both perturbation theory and numerical simulations find that the induced vison bandwidth by Zeeman terms is orders of magnitude smaller than the Majorana bandwidth, i.e. the effective vison mass $(m_v) \gg$ fermion mass $(m_f)$ \cite{Aprem21,Baskaran2023,Jin23} (see also supplement \cite{supp} for slow spreading of vison correlations). Thus, it allows for a quasi-static approximation of $\ket{D}$, by which the ground state sector of Eq.~\eqref{eq:Ham} at moderate $h$ can be approximated by a superposition of flux sectors and fermions:
\begin{equation} \label{eq:wavefunc}
    \ket{0} \sim \sum_{D} w_D(h) \ket{M_D}\otimes \ket{D},~~ \sum_{D} \abs{w_D(h)}^2 = 1
\end{equation}
where $w_D(h)$ is a coefficient associated with a quasi-static random field $\{D_{jl}\}$ [bottom of Fig.~\ref{fig:gapless}(a)]; and the summation is over an ensemble of $\{D_{jl}\}$, chosen to be consistent with the flux filling factor $\nu$ at intermediate field \cite{penghao24}. Typical disordered flux configurations have also been previously analyzed by static DMRG computations \cite{Baskaran2023}.

The approximation of Eq.~\eqref{eq:wavefunc} immediately evokes two aspects: 
(i) it is formally similar to the  setup of disorder-free localization, where the effective randomness originates from superposition \cite{Cirac05,Knolle17}; but here no auxiliary system is needed, since fractionalization itself provides the coherent disorder intrinsically. 
Within this Ansatz fermions are found to retain a system size-independent inverse-participation ratio (IPR) across the entire bandwidth, see Fig.~\ref{fig:gapless}(b). For the considered parameter it converges to around $\rm IPR \sim 0.1$, consistent with the estimated $\xi \sim 8 \sim \rm IPR^{-1}$ from our quench dynamics simulation, see supplemental materials \cite{supp}. 
For a small but finite $m_f/ m_v$ present in our model, visons are slow but not static and the fermions exhibit localization on transient times. The key point here is that the system in the ground state, and TL from fractionalization does neither require the preparation of a highly non-equilibrium initial state nor high temperatures~\cite{Heyl21,Nasu15}. 
We considered the ground state at finite fields, but a similar effect is expected to survive at  low temperatures; while at temperatures much higher than the microscopic scales effective fractionalization ceases to exist and, instead, conventional heat transport is expected~\cite{Brenig21,Aprem21}.

(ii) A concomitant phenomenon of the transient localization is its gapless spectrum. Each state $\ket{M_{D}}$ in the sum of Eq.~\eqref{eq:wavefunc} represents a class-D topological $p$-wave superconductor whose fermions propagate in a background of random $Z_{2}$ fluxes.  This setting produces a \emph{gapless} energy spectrum of a Majorana metal \cite{Senthil2000,Huse2012}, consistent with our MPS results in Fig.~\ref{fig:lattice}(d) and Fig.~\ref{fig:blurb}(a).  
We emphasize that, unlike a conventional “classical” Majorana metal, where disorder is extrinsic, i.e. either quenched or thermally activated \cite{Nasu15,Knollethermal}, the effective randomness in Eq.~\eqref{eq:Ham} emerges intrinsically from the fractionalization of its own degrees of freedom at low temperature, where one species of the fractionalized particles localizes the other.
A consequence of the Ansatz with visons treated as quasi–static, Eq.~\eqref{eq:wavefunc}, is that within this Ansatz the spin spectrum coincides with the averaged Majorana dynamics over flux sectors, see supplement \cite{supp}. 

{\it \tum {Fermion and spin spectral functions.-- }}
Now we show that the Majorana spectral function evaluated with Eq.~\eqref{eq:wavefunc} reproduces key features of the gapless spin structure factor obtained from our MPS simulations.
Consider randomly flipping flux pairs by changing the sign of $D_{jl}$ on each vertical bonds, illustrated by the red arrows in the bottom row of Fig.~\ref{fig:gapless}(a), with a probability $p$, which determines the flux density $\nu$.  At $p=0$, one recovers the gapped Majorana band for a $\pi$-flux ladder.
The emergence of gapless Majorana modes with increasing $p$ is captured by the ensemble-averaged translational symmetry of $\{M_D\}$ in Eq.~\eqref{eq:wavefunc}. 
For $0<p\ll1$, local in-gap Majorana resonances on flipped flux pairs have average separations $d\gg\xi$, so that the in‐gap states remain sharp and degenerate [Fig.~\ref{fig:gapless}(c)]. As $p$ increases, higher-order resonances emerge, which hybridize as $d \sim \xi$, lifting their degeneracy and broadening their spectral weight, ultimately producing gapless Majorana modes at sufficiently large  $p$ [Fig.~\ref{fig:gapless}(d,e)]. For a maximally random flipping of fluxes (quasi-static visons) given by $p = 0.5$, these in-gap Majorana resonances are merging into a gapless continuum; see Fig.~\ref{fig:gapless}(f).

The Majorana spectrum $A(k, \omega)$ in Fig.~\ref{fig:gapless}(f) with a flux filling $\nu \sim \frac{1}{2}$, as an effective theory to Eq.~\eqref{eq:Ham} under moderate $h$, should be compared to the numerical spin spectrum in Fig.~\ref{fig:lattice}(d). A more detailed argument regarding their mutual correspondence is presented in the supplement \cite{supp}. 
A good consistency is revealed: both the numerical result Fig.~\ref{fig:lattice}(d) and the effective theory Fig.~\ref{fig:gapless}(f) show a low-energy gapless Majorana continuum due to coherent flux disorder; and a higher energy continuum due to the broadened bulk Majorana band, albeit with some deviations near the band edge around $\omega \sim 2$;    
and zero-frequency spectra extending to a finite area around $\pm \pi$, revealing the metallic nature of Majorana fermions. Such an effective disorder-induced Majorana metal hosts a non-vanishing density of states, expected to result in a linear in temperature specific heat; it thus is expected to account for several features of field-induced QSL candidates and numerical simulations of extended Kitaev models, where the putative QSL phase appears to be gapless in specific-heat measurements \cite{Han21,Wei2025,zhou2025}.

\begin{figure}[t]
    \centering
    \includegraphics[width=\linewidth]{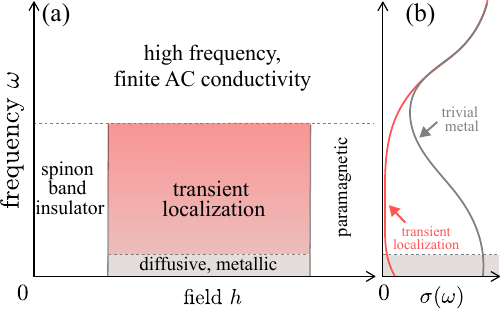}
    \caption{Mechanism of transient localization by fractionalization. (a) Schematic phase diagram of the Kitaev ladder model in a magnetic field. Transient localization can be observed at low energy scales (pink shaded) above the universal ultralow frequency regime (gray shaded). (b) Dynamical conductivity in the intermediate field regime. In contrast to a trivial metal which exhibits a strong Lorentzian peak as $\omega \to 0$ (gray solid line), due to transient localization from fractionalization the ac heat conductivity is suppressed at intermediate-to-low frequencies (red solid line). However, at ultralow frequencies, a small but finite conductivity will be recovered, as the vison mass is finite. }
    \label{fig:sum}
\end{figure}

{\it \tum {Concluding remarks.--} }
Using a Kitaev‐ladder model of spins which fractionalize into Majorana fermions (spinons) and fluxes (visons), we showed that gapless fermions and suppressed transport can coexist in a translation-invariant phase without extrinsic disorder. The essential mechanism is a form of coherent randomness that arises \emph{intrinsically} from fractionalizing the system’s own local degrees of freedom: the heavy vison fluctuations around the ground state act as a self‐generated, quasi‐static disorder that localizes the Majorana fermions. A minimal Ansatz, Eq.~\eqref{eq:wavefunc}, which treats the ground state as a superposition of static flux sectors, can account for the localization physics. The Ansatz lies beyond the reach of self‐consistent mean fields, see supplemental materials \cite{supp}, as self-averaged order parameters cannot capture the nonlocal interference responsible for localization \cite{Pastor2003}. At low but finite temperatures localization will persist as more flux sectors within a range $k_B T$ contribute. At zero temperature the finite-frequency transport behavior (which we focus on) can similarly be understood within this framework (pink shaded area in Fig.~\ref{fig:sum}(a)). However, our system is ultimately non-integrable and visons have a large but finite mass. As a result, tunneling between vison configurations is probed at asymptotically low frequencies or temperatures; thus, ultimately the system shows diffusive heat conductivity at ultra low scales~\cite{Schiulaz2014, DeRoeck2014, Papic2015, Moore16, DarkwahOppong2020} with a suppressed Drude weight \cite{Fratini21, Hadi24} (red solid line). This should be contrasted with a large Drude peak of a conventional diffusive gapless system (gray solid line). Accessing the universal regime experimentally will demand reaching exceedingly low frequencies and realizing ultra pure samples.

We anticipate the transient localization arising from intrinsic quantum coherent disorder to be a key phenomenon for interpreting gapless quantum spin liquids. Crucially, it shows that the standard parton mean-field approaches which assume static translational flux configurations are insufficent. For future work, it will be pertinent to compute the temperature dependence of thermodynamic quantities, consider extrinsic disorder effects at ultralow energy scales, explore the mechanism of transient localization with parton constructions of gapless quantum spin liquids, and compare the effective theory with experimental findings. Our mechanism of transient localization from fractionalization demonstrates that the suppressed dynamical response at intermediate frequencies or thermal transport at intermediate temperatures can be an intrinsic property of fractionalization in gapless quantum spin liquids.

\acknowledgments{
{\it \tum{Acknowledgments.-- }}
We are grateful to W. Kadow, X. Feng, K. B. Yogendra, N. Trivedi and F. Pollmann for comments and discussions.   
S.F. J.K. and M.K. acknowledge support from the Deutsche Forschungsgemeinschaft (DFG, German Research Foundation) under Germany’s Excellence Strategy--EXC--2111--390814868, TRR 360 – 492547816 and DFG grants No. KN1254/1-2, KN1254/2-1, the Imperial-TUM flagship partnership, the European Research Council (ERC) under the European Union’s Horizon 2020 research and innovation programme (grant agreement No. 851161), the European Union (grant agreement No 101169765), as well as the Munich Quantum Valley, which is supported by the Bavarian state government with funds from the Hightech Agenda Bayern Plus. P.Z acknowledges support from U.S. National Science Foundation, NSF-MRSEC under Award No. DMR-2011876.
}

{\it \tum{Data availability.-- }}
Data, data analysis, and simulation codes are available upon reasonable request on Zenodo~\cite{zenodo}.

\bibliography{biblio}

\clearpage
\begin{widetext}
\begin{center}
\textbf{\large Supplementary material for ``\textbf{\ourtitle}"}

\medskip
Shi Feng,$^{1,2}$ 
Penghao Zhu,$^{3}$ Johannes Knolle,$^{1,2,4}$, and Michael Knap$^{1,2}$

\medskip
{\it \small ${}^1$Technical University of Munich, TUM School of Natural Sciences, Physics Department, 85748 Garching, Germany}

{\it \small ${}^2$Munich Center for Quantum Science and Technology (MCQST), Schellingstr. 4, 80799 M{\"u}nchen, Germany}

{\it \small ${}^3$Department of Physics, The Ohio State University, Columbus, Ohio 43210, United States}

{\it \small ${}^4$Blackett Laboratory, Imperial College London, London SW7 2AZ, United Kingdom}

\end{center}
\end{widetext}

\setcounter{equation}{0}
\setcounter{figure}{0}
\setcounter{table}{0}
\setcounter{page}{1}
\setcounter{section}{0}
\makeatletter
\renewcommand{\theequation}{S\arabic{equation}}
\renewcommand{\thefigure}{S\arabic{figure}}
\renewcommand{\bibnumfmt}[1]{[S#1]}
\setcounter{secnumdepth}{3}  
\makeatletter
\def\thesection{S\arabic{section}}
\makeatother
\let\addcontentsline\oldaddcontentsline
\tableofcontents

\section{The Integrable Limit and the Ground-State Sector}
In this section we present the ground-state sector and the solution to the integrable ladder model at $h=0$:
\begin{equation}
    H_{\rm ladder} = \sum_{j + l \in \rm even} \left(X_{j}^l X_{j+1}^l + Y_{j-1}^l Y_{j}^l + Z_{j}^l Z_{j}^{l+1} \right) \label{eq:ladder}
\end{equation}
where the subscript $j$ is the horizontal index along the ladder, and the superscript $l$ a binary index labeling the lower ($l=0$) or upper ($l=1$) leg. 
It can be reduced to the form of fermion bilinear either by Kitaev's four-Majorana decomposition \cite{kitaev2006anyons} or a Jordan-Wigner transformation \cite{Xiang07}. The former method adopt a local decomposition of spin into fermions which doubles the Hilbert space, thus it requires a fermion parity projection to get the physical wavefunction; while the latter retains the same size of the physical Hilbert space using a non-local Jordan-Wigner-type fermionization. 
\begin{figure}[b]
    \centering
    \includegraphics[width=\linewidth]{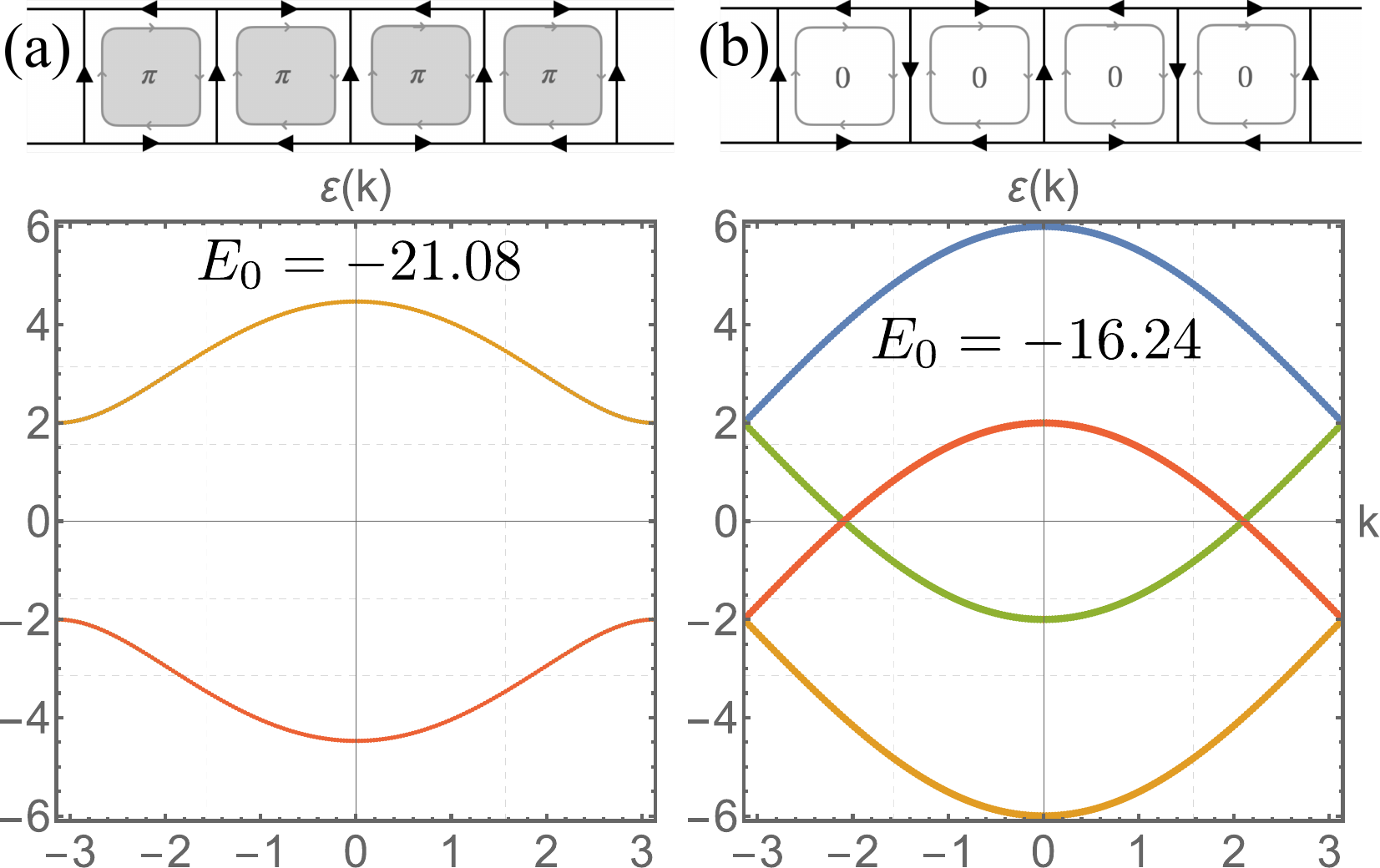}
    \caption{(a) The ground state sector with a uniform $\pi$-flux configurations and gapped Majorana bands conditioned thereof. (b) Majorana bands under flux-free condition. Data are obtained by diagonalizing the $800$‐site tight‐binding Majorana Hamiltonian with $Z_2$ gauge fields indicated by black arrows. The ground‐state energy $E_0$ in the $\pi$‐flux configuration lies significantly below that of the flux‐free case. }
    \label{fig:bands}
\end{figure}
We here present the exact solution and the $Z_2$ gauge field for $H = H_0$ in the main text. Due to the simple ladder geometry, it is more convenient to solve the model by using Jordan-Wigner transformation \cite{Xiang07,Chen2008}: 
\begin{align}
    P_{j}^0 &= 2 \left(\prod_{j<j'} Z_{j'}^0\right) \left(a_j^0\right)^\dagger,\\
    M_j^0 &= 2 a_j^0 \left(\prod_{j<j'} Z_{j'}^0\right), \\
    P_{j}^1 &= 2 \left( \prod_{j'} Z_{j'}^0 \right) \left(\prod_{j<j'} Z_{j'}^1\right) \left(a_j^1\right)^\dagger, \\
    M_j^1 &= 2 a_j^1 \left( \prod_{j'} Z_{j'}^0 \right) \left(\prod_{j<j'} Z_{j'}^1\right), \\
    Z_j^l &= 2 (a_j^l)^\dagger a_j^l - 1,~ \forall \,l \in \{0,1\}
\end{align}
where $P = X + iY$, $M = X-iY$, $a$ and $a^\dagger$ are canonical fermion operators, and the super-(sub-) scripts inherit the same convention from the Pauli matrices in Eq.~\eqref{eq:ladder}.
This allows us to Majorana-fermionize the zero-field Hamiltonian without introducing redundancy into
\begin{equation} \label{eq:ladderf}
    H_{\rm ladder} = i \sum_{j+l \in \rm even} \left( c_{j+1}^l c_j^l + c_{j-1}^l c_j^l + D_{jl} c_j^{l+1} c_{j}^l\right)
\end{equation}
where $D_{jl}$ are good quantum numbers taking values $\pm 1$. 
In the spin representation, the binary Majorana fluxes are given by square faces
\begin{equation} \label{eq:fluxops}
    F_a = X^0_0 Y^1_0 X^0_1 Y^1_1,\;\;F_b = Y^0_1 X^1_1 Y^0_2 X^1_2
\end{equation}
modulo translations along the horizontal subscript $j$. 
The ground state flux of each square plaquette $p$ in  Eq.~\eqref{eq:ladderf}, taking the clock-wise convention, is
\begin{equation}
    \phi_{\rm gs}(p) = \prod_{\langle jk\rangle\in p}(-i\,u_{jk}) = - F_{a(b)}
\end{equation}
where the anti-symmetric $u_{jk}$ is given by the hopping amplitude in Eq.~\eqref{eq:ladderf}, i.e. $\pm 1$ on horizontal links or $D_{jl}$ on vertical links. 
The ground-state fermion is given by a uniform filling of $\pi$ fluxes corresponding to $\expval{F_a} = \expval{F_b} = 1$. We note that the flux pattern for the ground state sector is different from the honeycomb model, which is flux-free in the ground state. One might relate the Kitaev ladder model to a honeycomb model with $L_y = 2$ under periodic boundary condition, but the flux filling at the ground states are distinct. 
This is because the elementary plaquettes in the ladder are no longer hexagons but squares faces, thus the ground-state sector of the Kitaev ladder are given by a uniform $\pi$ flux filling as described by Lieb's theorem \cite{Lieb1994}. In contrast, the flux-free condition would result in a higher-energy sector and an incorrect Majorana band structure at lowest energies.  This is shown in Fig.~\ref{fig:bands}.

\section{Inverse Participation Ratio}
In our Kitaev‐ladder model, all disorder arises from magnetic‐field–induced fluctuations of the $Z_2$ gauge fluxes (visons). A finite field $h$ breaks the exact local conservations and endows each vertical‐bond gauge variable $D_{jk}=\pm1$ with a small but nonzero bandwidth, that is, a finite vison mass $m_v$, so that the true ground state becomes a coherent superposition of random flux configurations with filling factor $\nu$. In the regime $m_f\ll m_v$, these visons are slow on the timescales of Majorana dynamics and can be treated as quasi‐static, translation‐invariant disorder. We emphasize that $m_v$ is large but finite, so this static‐flux approximation strictly holds up to intermediate times only.

\begin{figure}[t]
    \centering
    \includegraphics[width=\linewidth]{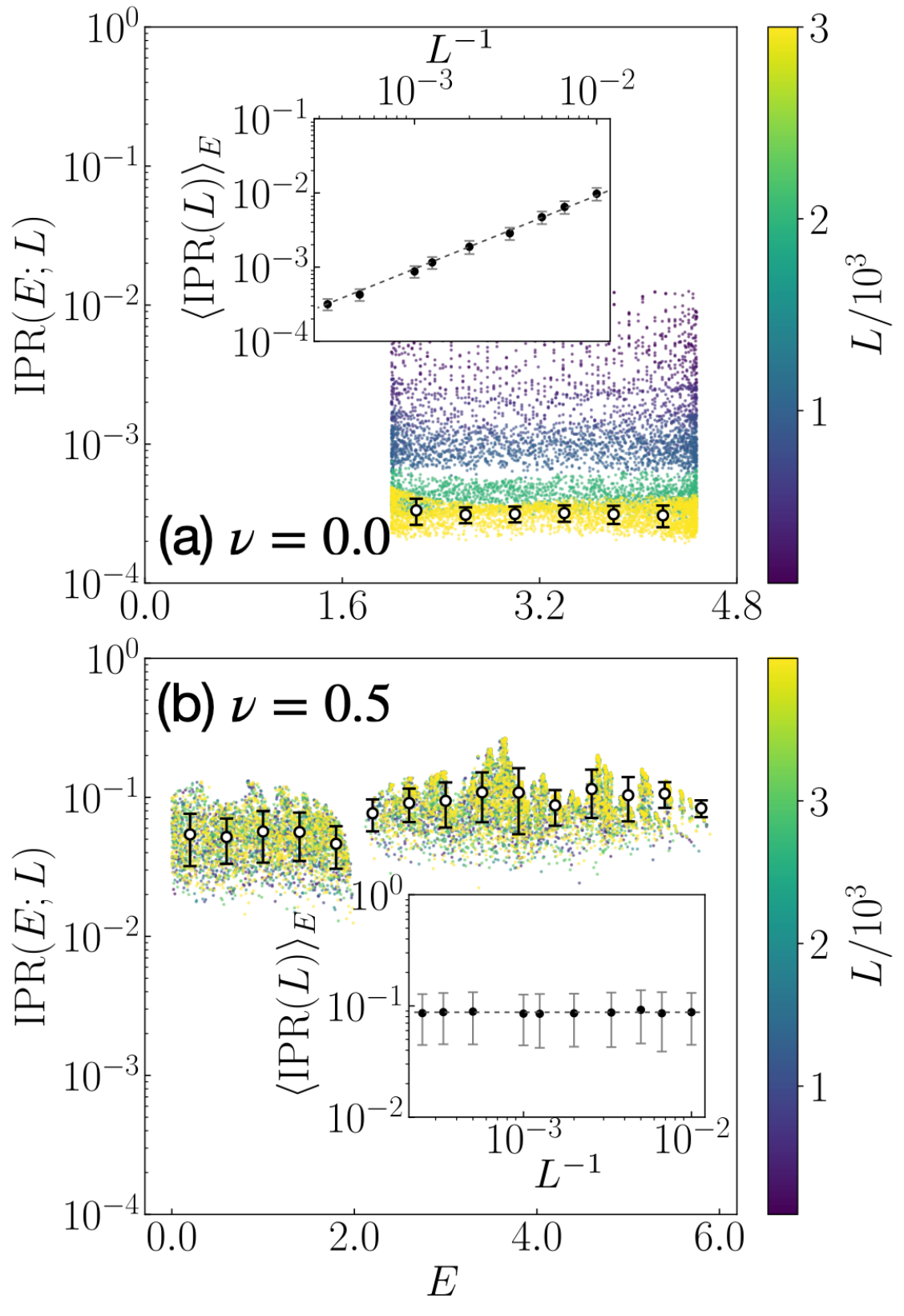}
    \caption{Different measure of IPRs versus energy for vison fillings $\nu=0.0$ (a) and $0.5$ (b), respectively, at various system sizes; colored markers indicate ${\rm IPR}(n,L)$ for different system sizes from $L=500$ to $L=4000$; hollow circles with error bars denote ${\rm IPR}(E,L)$ for $L=4000$. Insets: finite-size scaling of the bandwidth-averaged $\bigl\langle \mathrm{IPR}(L)\bigr\rangle_E$. Error bars denote the standard deviation from the average value. }
    \label{fig:ipr}
\end{figure}

To quantify the effective localization, we compute the energy‐resolved inverse participation ratio (IPR) for eigenstates across the full bandwidth
\begin{equation}
    \mathrm{IPR}(n,L)=|\psi_n(i)|^4,
\end{equation}
its binned average
\begin{equation}
    \mathrm{IPR}(E,L)=\frac{1}{N(E)}\sum_{n:E_n\approx E}\sum_{i}|\psi_n(i)|^4,
\end{equation}
and the averaged ${\rm IPR}(L)$ over all energies
\begin{equation}
    \bigl\langle \mathrm{IPR}(L)\bigr\rangle_E
    \;=\;\frac{1}{N_{\rm tot}}\sum_{n=1}^{N_{\rm tot}}\mathrm{IPR}(n,L)
\end{equation}
where $\psi_n(i)$ is the $n$th normalized Majorana eigenmode on site $i$, $E_n$ its energy, $N(E)$ the number of states in the bin, and $N_{\rm tot}$ is the total number of states within the full bandwidth.  Figure~\ref{fig:ipr}(a) shows $\mathrm{IPR}(E,L)$ for the clean KSL ($\nu=0.0$) at $L=500$–$4000$ (purple–yellow).  The white‐circle markers (at $L=4000$) trace the lower envelope of the data cloud, and the inset in Fig.~\ref{fig:ipr}(a) confirms that the bandwidth‐averaged IPR decays as $\langle\mathrm{IPR}(L)\rangle_E\propto L^{-\tau}$, demonstrating that all Majorana modes are extended as $L\to\infty$.

By contrast, Fig.~\ref{fig:ipr}(b) presents $\mathrm{IPR}(E,L)$ for a maximally random flux filling ($\nu=0.5$).  Here the IPR values cluster around $10^{-2}-10^{-1}$ with no systematic drift as $L$ grows, and the inset shows $\langle\mathrm{IPR}(L)\rangle_E$ remains finite and essentially constant with system size.  This unambiguous size‐independence across the full bandwidth signals that the coherent vison disorder localizes all Majorana eigenmodes, realizing disorder‐free localization in a translation‐invariant, zero‐temperature ground state. Moreover, one can show that the localization length $\xi$ is proportional to $\text{IPR}^{-1/d}$ in $d$ spatial dimensions. Specifically, consider an exponentially localized state with wavefunction $|\psi(\mathbf{r})|^2=e^{-2r/\xi}f(\Omega_{d-1})$ up to a normalization factor, where $\Omega_{d-1}$ is the solid angle on the d-sphere. and $f(\Omega_{d-1})$ captures the angular dependence. The IPR is then calculated as
\begin{equation}
\label{eq:IPRunnormalized}
\begin{aligned}
&\text{IPR}=\frac{\int _{0}^{+\infty}d\mathbf{r}|\psi(\mathbf{r})|^4}{\left[\int _{0}^{+\infty}d\mathbf{r}|\psi(\mathbf{r})|^2\right]^2}
\\
&=\frac{\int_{0}^{+\infty} dr r^{d-1}\exp(-4r/\xi) \int d\Omega_{d-1} f^2(\Omega_{d-1}) }{\left[\int_{0}^{\infty}dr r^{d-1}\exp(-2r/\xi)\int d\Omega_{d-1}f(\Omega_{d-1})\right]^2}.
\end{aligned}
\end{equation}
Using $\int_{0}^{+\infty} dr r^{d-1} \exp(-r/a)=a^{d}\Gamma(d)$ where $\Gamma(d)=(d-1)\Gamma(d-1)$ is the gamma function, we obtain
\begin{equation}
\text{IPR}=\frac{(\xi/4)^{d}\Gamma(d)\int d\Omega_{d-1}f^2(\Omega_{d-1})}{(\xi/2)^{2d}(\Gamma(d))^2(\int d\Omega_{d-1}f(\Omega_{d-1}))^2} \propto \xi^{-d}.
\end{equation}
Therefore, $\xi \propto \text{IPR}^{-1/d}$. In our case with $d=1$, we find $\xi \propto \text{IPR}^{-1}$, which is consistent with our numerical results discussed in the main text.

Hence, in the quasi‐static limit $m_f/m_v \to 0$, or equivalently, for the translation‐invariant flux disorder of Eq.~(3), all Majorana eigenstates become localized within a length scale $\xi$ and the local energy will not spread beyond $\xi$ following a local quench. For small but finite $m_f/m_v$ in the transient DFL regime, this localization persists over observable transient timescales, as demonstrated by the strongly suppressed quench dynamics and vanishing energy transport presented in the main text.

\section{Connecting Spin Structure Factor and Majorana Spectrum}
In the limit $m_f/m_v \to 0$, the spin dynamics would be governed by the Majorana fermions. 
The quasi-static ansatz in the main text
\begin{equation} \label{eq:appwavef}
    \ket{0} \sim \sum_{D} w_D(h) \ket{M_D}\otimes \ket{D}
\end{equation}
is thus validated by showing the consistency between the gapless Majorana spectral function using the quasi-static approximation and the spin spectral function obtained by MPS algorithm.  
Note that in this variational ansatz for $h\neq0$, the state $\lvert M_D\rangle$ in principle need not be the exact Fermi sea for the fixed gauge configuration $D$; in presence of the non-commuting Zeeman term, it is also reasonable to assume, in general, that $M_D$ also includes those fermion states from nearby flux sectors $D\pm\delta D$ in keeping with $\nu \sim 0.5$. 
Without detailed knowledge of those $\lvert M_D\rangle$ one cannot derive an exact result.  However, under the quasi-static vison approximation, and given that the single-fermion states remain strongly localized over the entire bandwidth for a finite window of $(D\pm\delta D)|_{\nu \sim 0.5}$, i.e. insensitive to a finite variation in $\delta D$, this construction still provides a reliable semi-quantitative estimate.

The dynamic spin correlation can be rewritten in terms of fermion and flux basis. For simplicity, we present the correlation between $Z_i$ and $Z_j$ as an example.
\begin{equation}
\begin{aligned}
\label{eq:spincorrproof}
    &\expval{0|Z_i(t) Z_j|0} \\ &=\sum_{D,D^{\prime}}w^{\star}_{D^{\prime}}w_{D}e^{iE_{0}t}\bra{D^{\prime}\,M_{D^{\prime}}}Z_{i}e^{-i Ht}Z_{j}\ket{M_{D}\,D}
    \\ &=\sum_{D,D^{\prime}}w^{\star}_{D^{\prime}} w_{D}e^{iE_{0}t}\bra{D_i'\,M_{D^{\prime}}}c_{i}e^{-i H t}c_{j}\ket{M_{D}\,D_j}
    \\ &\simeq \sum_{D,D^{\prime}}w^{\star}_{D^{\prime}} w_{D}e^{iE_{0}t}\delta_{D^{\prime}_{i},D^{\phantom{\prime}}_{j}}\bra{M_{D^{\prime}}}c_{i}e^{-i H_M^j t}c_{j}\ket{M_{D}}
    \\ &= \sum_{D} w^{\star}_{D_{ij}} w_{D}e^{iE_{0}t}\bra{M_{D_{ij}}}c_{i}e^{-i H_M^j t}c_{j}\ket{M_{D}}
\end{aligned}
\end{equation} 
where $D_j$ ($D_{ij}$) denotes the gauge configuration with $j$-th ($i$- and $j$-th) vertical bond(s) flipped, and in the third equality we assumed that in the quasi-static limit of $m_f \ll m_v$ the time evolution of fluxes can be ignored in transient timescales, thus $\expval{D'_i|H|D_j} \simeq \delta_{D_i',D_j} H_{M}^j$, with $H_{M}^j$ the tight-binding Majorana model conditioned on the gauge sector $D_j$. Note this is different from the Kitaev model at $h=0$ whereby only the nearest neighbors contribute -- the extensive superposition of gauge configurations in Eq.~\eqref{eq:appwavef} introduces long-range spin correlations. 
The evaluation of Eq.~\eqref{eq:spincorrproof} can be divided into two parts: 
(1) When $i,j$ are nearest neighbors, $Z_2$ link satisfies $D_{ij} = D$, hence the matrix elements in Eq.~\eqref{eq:spincorrproof} become $\expval*{M_D| c_i e^{-iH_M^j} c_j| M_D}$. This is similar to the nearest neighbor contribution in the integrable limit \cite{Knolle2014}. 
(2) When $i$ and $j$ lie farther apart, however, the matrix elements of
$\bra{M_{D_{ij}}}c_{i}\,e^{-iH_M^j t}\,c_{j}\ket{M_{D}}$
are challenging to evaluate exactly without an explicit form for $\ket{M_{D_{ij}}}$.

These are quench problems in that $\ket*{M_D}$ or $\ket*{M_{D_{ij}}}$ conditioned on $D$ or $D_{ij}$ is not an eigenstate of $H_M^j$ conditioned on $D_j$. Its exact evaluation may be done via the evaluation of Pfaffians \cite{Knolle2014}. Here we take a simplified approach, namely the adiabatic approximation which captures the majority of spectral information at a low cost \cite{Knolle2014,Knolle15}. Under the adiabatic approximation, one considers the scattering potential turned on at $t = -\infty$ instead of $t=0$.  Effectively, one replaces, for example, the state $\ket*{M_D}$ conditioned on gauge field $\ket{D}$ with the state $\ket*{M_{D_j}}$ conditioned on the gauge field $\ket{D_j}$ that has the link on $j$ flipped.
The matrix elements in both cases (1) and (2) can hence be computed in the diagonal form $\expval*{M_{D_j}| c_i e^{-iH_M^j} c_j| M_{D_j}}$, where $\ket{M_{D_j}}$ is the Fermi sea conditioned on the flux sector $\ket{D_j}$.  Assuming a near-uniform weight distribution in $w_D$, Eq.~\eqref{eq:spincorrproof} becomes the disorder-averaged Lehmann spectral function:
\begin{equation} \label{eq:spec}
\begin{split}
    \expval{Z_i(t) Z_j} &\simeq \sum_{D} \abs{w_{D}}^2 e^{iE_{0}t}\bra*{M_{D_j}}c_{i}e^{-i H_M^j t}c_{j}\ket*{M_{D_j}}
        \\
    &= \sum_{D}\abs{w_{D}}^2\bra{M_{D}}c_{i}(t)c_{j}\ket{M_{D}}
\end{split}
\end{equation}
We point out that since the fermion subsystem has a finite DOS at zero energy under vison fluctuation, the adiabatic approximation would ignore the X-ray-edge scaling factor $\omega^{\alpha - 1}$ \cite{DOMINICIS}, which underestimates (overestimates) the intensity as $\omega \to 0$ ($\omega \gg 0$). Indeed, this can be seen in the stronger gapless intensities in the numerical results [Fig.~1(d) of the main text] in comparison to the effective spectral function [Fig.~3(f)].  
Nevertheless, as a semi-quantitative estimate for the spectral function, it still captures the most important feature of relevance, i.e. the presence and the position of gapless spectra.  
Equation~\eqref{eq:spec} is the average of the Majorana spectral functions over a random ensemble of gauge configurations with a fixed flux filling factor $\nu$ suggested by $\expval*{F_{a(b)}} \sim 0$ under moderate magnetic field. Correlations between $X$ and $Y$, etc. can be done by the same token. In the next section, we show how the momentum-resolved average is obtained for Fig.~3(c-f) in the main text.

\begin{figure*}[t]
    \centering
    \includegraphics[width=\linewidth]{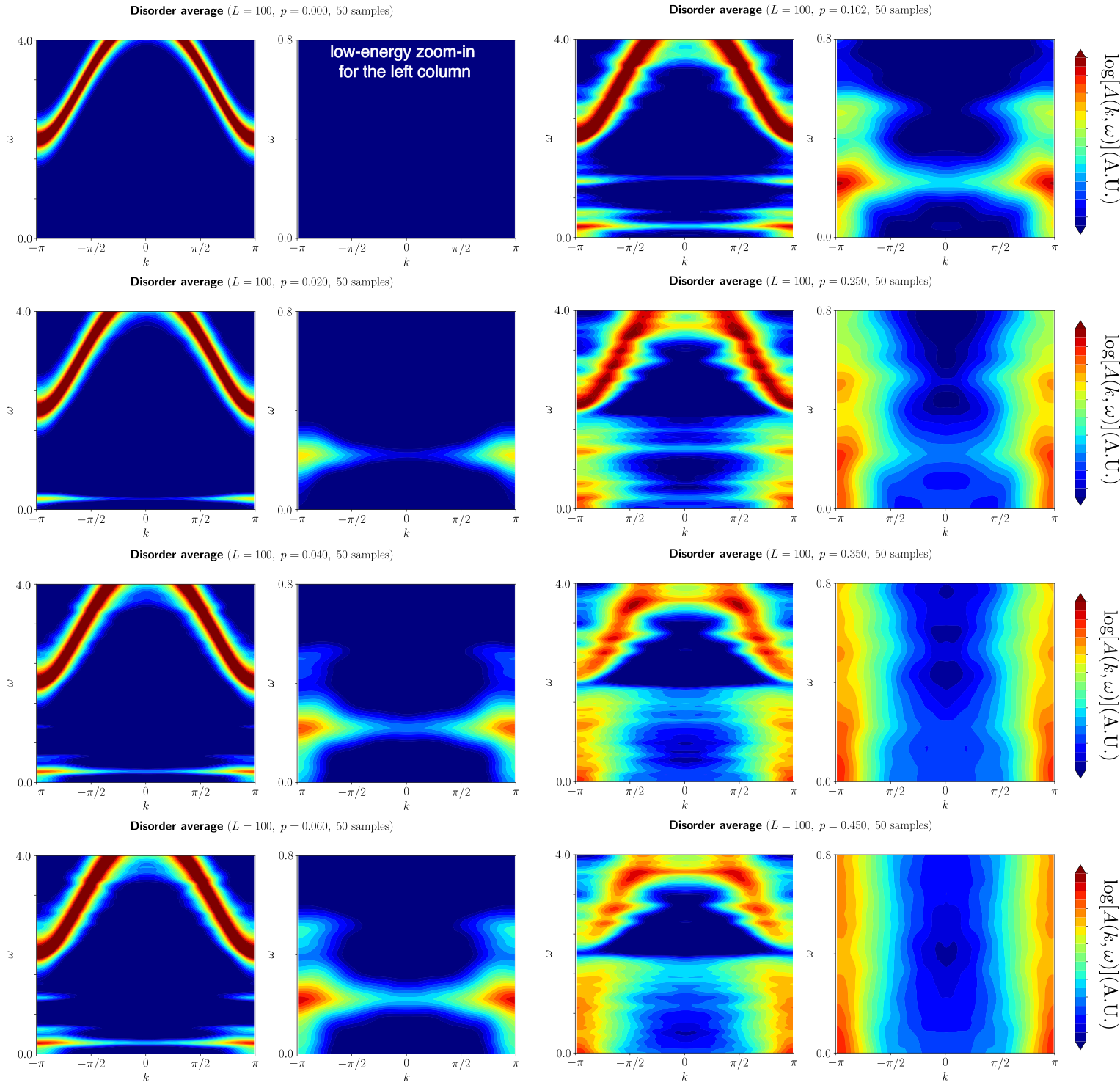}
    \caption{The average-momentum-resolved Majorana spectral functions. This showcases the gradual emergence of gapless Majorana modes near BZ boundary as a function of increasing flux density. Data obtained by the disorder average given in Eq.~\eqref{eq:disav}.}
    \label{fig:Akw_avgs}
\end{figure*}

\section{ Disorder-Average Spectral Functions}
Equation~\eqref{eq:spec} states that the spin spectral function is given by the disorder-averaged Majorana spectral function $A(r_i, r_j, \omega) =\sum_{D}\abs{w_{D}}^2\bra{M_{D}}c_{i}(t)c_{j}\ket{M_{D}}$. With disorder in each summand in Eq.~\eqref{eq:spec}, the translation symmetry is broken, however, momentum-resolved spectrum $A(k, \omega)$ can be recovered by considering the average translation symmetry. In this section we derive $A(k, \omega)$ by averaging Majorana spectral function over the center-of-mass position \cite{penghao24}, which is used for producing Fig.~3(a-c) in the main text. 

It is convenient to introduce four sub-lattice sites per unit cell and to label the real-space orbitals by $|r,s\rangle$ with $r = 0,\dots ,L-1, s = 0,1,2,3$
being the unit-cell index and sublattice index respectively. 
An eigenstate $\ket{n}$ of an \emph{disordered} Hamiltonian $H_{\text{dis}}$ is 
\begin{equation} \label{eq:diseigen}
  |n\rangle
    = \sum_{r,s} \phi_n(r,s)\,|r,s\rangle,\,
  \phi_n(r,s)=\langle r,s|n\rangle .
\end{equation}
For the translational invariant ladder we diagonalize the $4\times4$
Bloch Hamiltonian $h(k)$ at momenta
$k_m = 2\pi m/L\;(m=0,\dots ,L-1)$ and obtain four orthonormal
eigenvectors $u_{\alpha k}\;(\alpha=1,\dots ,4)$.
The corresponding Bloch state is
\begin{equation}
  |s,k\rangle
    = \frac{1}{\sqrt L}
      \sum_{r=0}^{L-1}\sum_{s=0}^{3}
      u_{s,k}\,e^{ikr}\,|r,s\rangle ,
\end{equation}
with real-space Bloch wavefunction  $\psi_{s, k}(r) = \frac{u_{s, k}}{\sqrt L}\,e^{ikr}$. 
The $4L$ Bloch states form a complete orthonormal set $\sum_{\alpha=1}^{4}\sum_{k\in\text{BZ}}|s,k\rangle\langle s,k|
= \openone_{4L}$.
Inserting this resolution of identity into Eq.~\eqref{eq:diseigen} expands a generic Majorana state under quasi-static flux disorder in the translation-invariant Bloch basis: $ |n\rangle = \sum_{s=0}^{3}\sum_{k\in\text{BZ}} c^{n}_{s, k}\,|s,k\rangle$, 
where the coefficient $c^{n}_{s, k} = \langle s,k|n\rangle$ is given by
\begin{equation} \label{eq:csnk}
    c^{n}_{s, k} = \sum_{r=0}^{L-1}\sum_{s=0}^{3} \psi^{*}_{\alpha k}(r,s)\;\phi_n(r,s) 
\end{equation}
whose average over an ensemble of $D$ contains all the information in the Majorana spectral function $A(k, \omega)$. This can be seen explicitly by averaging each
\begin{equation} \label{eq:disav}
    A(r_i, r_j, \omega) = \sum_n \tr_s \phi_n^*(r_i,s) \phi_n(r_j,s) \delta(\omega - E_n)
\end{equation}
over the center-of-mass position $R = (r_i + r_j) / 2$, i.e. $\expval{A(r_i, r_j, \omega)}_R = \frac{1}{L}\sum_R A(r, R, \omega)$,  which gives
\begin{equation}
\expval{A(r_i, r_j, \omega)}_R 
= \frac{1}{L} \sum_{n,s,k} e^{ikr} \abs*{c_{s,k}^n}^2 \delta(\omega - E_n)
\end{equation}
Hence averaging the norm square of Eq.~\eqref{eq:csnk} suffice to give the average-momentum-resolved Majorana spectral function $A(k, \omega)$ as a good approximation of spin spectral function [Eq.~\eqref{eq:spec}], as is shown by Fig.~3(d) in the main text. 
In Fig.~\ref{fig:Akw_avgs}, we present additional results for $A(k,\omega)$ under different flux densities, showcasing the gradual emergence of gapless Majorana modes.

\section{Discrete Continuity}
In this section we derive Eq.~(4) in the main text by the discrete continuity equation, and define the energy density and energy current operators used therein.  
\subsection{Relation between correlation functions}
In the discrete system, the discrete version of the continuity equation for local energy density $\mathcal{E}_r$ is
\begin{equation} \label{eq:dcon}
    \frac{d}{dt} \mathcal{E}_r(t) + \left[J_{r, r+\delta}(t) - J_{r-\delta, r}(t) \right]= 0
\end{equation}
where $J_{r, r+\delta}$ denotes the local energy current from site $r$ to its nearest neighbor $r + \delta$. 
Applying the discrete Fourier transform in space time
\begin{align}
    \mathcal{E}(k, \omega) &= \sum_r \int dt\, e^{-ik r + i \omega t} \mathcal{E}_r(t),\\    J(k, \omega) &= \sum_{r} \int dt\, e^{-ik r + i\omega t} J_{r, r+\delta}(t),
\end{align}
we rewrite Eq.~\eqref{eq:dcon} into
\begin{equation}
    -i\omega \mathcal{E}(k, \omega) + \left( 1 - e^{-ik\delta} \right) J(k,\omega) = 0
\end{equation}
In the long wavelength limit, i.e., $1 - e^{-ik\delta} \approx ik\delta$, the conventional  continuity equation is obtained:
\begin{equation}
    -i\omega \mathcal{E}(k, \omega) + ik\delta J(k, \omega) = 0 .
\end{equation}
So that we can relate the long-wavelength current spectral function and energy spectral function by
\begin{equation}
    \expval{J(k, \omega) J(-k, \omega)} = \frac{\omega^2}{k^2 \delta^2} \expval{\mathcal{E}(k, \omega) n(-k, \omega)}
\end{equation}
However, since our goal is to inspect the spectrum across the whole Brillouin zone, we will keep the higher order terms in $k$. For a generic $k$, we have
\begin{equation}
    J(k, \omega) = \frac{\omega}{1 - \exp(-ik\delta)} \mathcal{E}(k, \omega)
\end{equation}
hence the correlation functions are related by
\begin{equation} \label{eq:appeejj}
    \expval{J(k, \omega) J(-k, \omega)} = \frac{\omega^2}{2 - 2\cos(k\delta)} \expval{\mathcal{E}(k, \omega) \mathcal{E}(-k, \omega)}
\end{equation}
that is the Eq.~(4) in the main text (with $\delta = 1$). This relation holds for all discrete time-independent systems. To evaluate these correlation functions in our system, in the following section we define energy and energy current operators in the Kitaev ladder model.

\subsection{Energy-density and energy-current operators}
\begin{figure}[t]
    \centering
    \includegraphics[width=\linewidth]{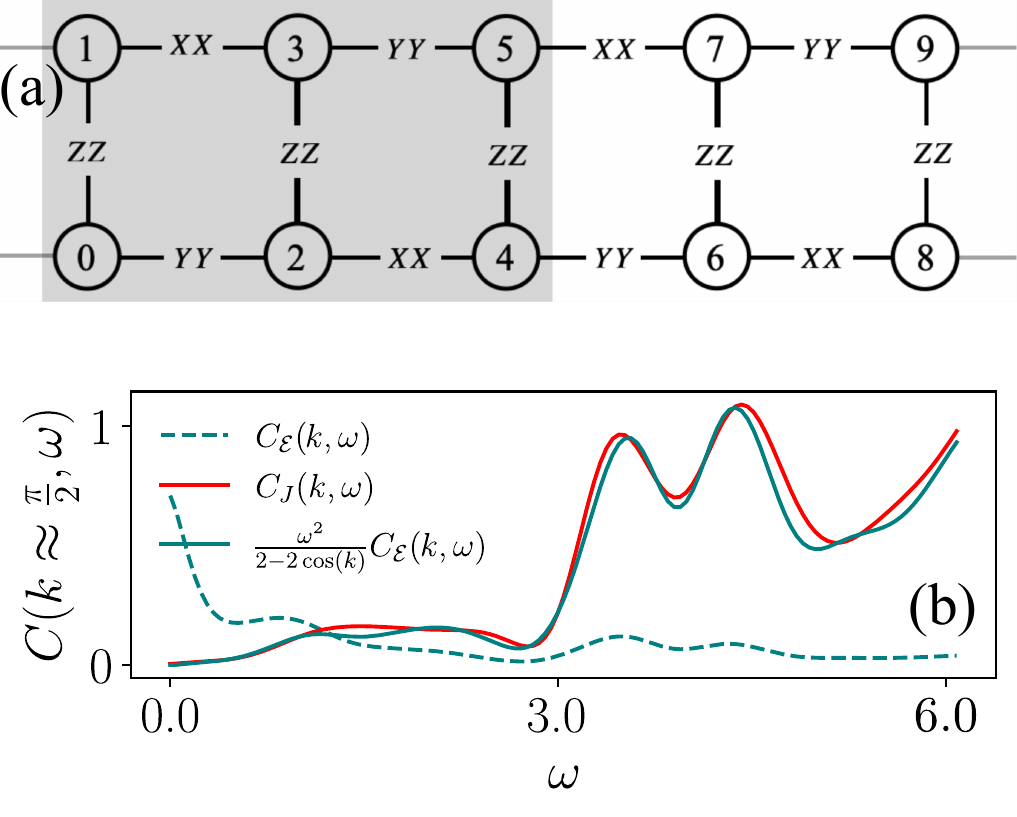}
    \caption{(a) The Kitaev ladder model with simplified indices, where the upper and the bottom rows are assigned odd and even indices respectively. The local energy density operator is defined within the shaded region. 
    (b) Consistency check by the comparison between the left- and right-hand side of Eq.~\eqref{eq:appeejj} at $k=\frac{\pi}{2}$. 
    }
    \label{fig:ladder2}
\end{figure}

For simplicity and clarity, we label sites on the ladder in the way shown in Fig.~\ref{fig:ladder2}(a) without using the superscript. The upper and the bottom rows are now assigned odd and even indices respectively. 
The definition of local energy density is not unique. Here we define energy density with a symmetric support under reflection with respect to $z$ bonds.  As is shown in Fig.~\ref{fig:ladder2}(a), we use the middle site at the bottom leg within the shaded region as the site of reference, and define the local energy $h$, thus its fluctuation $\mathcal{E} \equiv h - \expval{h}$, in the shaded region as
\begin{equation}
\begin{split} \label{eq:energydensity}
    h_{2} & = Z_2 Z_3 + X_1 X_3 + X_2 X_4 + Y_0 Y_2 + Y_3 Y_5 \\
    & + \frac{1}{2}(Z_0 Z_1 + Z_4 Z_5) \\
    &+ \mathbf{h}\cdot (\boldsymbol{\Pi}_2 + \boldsymbol{\Pi}_3) + \frac{\bf h}{2}\cdot (\boldsymbol{\Pi}_0 + \boldsymbol{\Pi}_1 + \boldsymbol{\Pi}_4 + \boldsymbol{\Pi}_5)
\end{split}
\end{equation}
modulo translations, 
where $\boldsymbol{\Pi} = (X, Y, Z)$, and the magnetic field ${\bf h}=(h_x, h_y, h_z)$; $h_{2}$ denote the local energy density operator whose site of reference is $2$, as shown in Fig.~\ref{fig:ladder2}(a). It is straightforward to see that this definition recovers the total Hamiltonian with $H = \sum_i h_i$.  
The energy current operator $J$ therefore can be defined by Heisenberg equation of motion and the aforementioned continuity equation:
\begin{align}
\begin{split} \label{eq:energycurrent}
    J_2 &= 2(X_2 Z_4 Y_6 - Y_3 Z_5 X_7) \\
    &+ (Z_4 Y_5 X_7 + Y_3 Z_4 X_5 - X_2 Y_4 Z_5 - X_4 Z_5 Y_6) \\
    &+ h(X_5 Y_3 - X_2 Y_4 + X_7 Y_5 - X_4 Y_6) \\
    &+ h[(X_2 + Y_6) Z_4 - (X_7 + Y_3) Z_5]
\end{split}
\end{align}
Equation~\eqref{eq:energycurrent} and Eq.~\eqref{eq:energydensity} are used in Eq.~(4) of the main text for the MPS time evolution.  

By these definitions we compute the dynamical autocorrelations in Eq.~(4) of the main text [c.f. Eq.~\eqref{eq:appeejj}] across the whole BZ. As is shown in Fig.~2 of the main text, 
$C_\mathcal{E}$ is gapless at $k_0 = \pm \frac{\pi}{2}$ where the signal diverges at small $\omega$ as $C_\mathcal{E}(k_0, \omega) \propto \omega^{-a}$ with $2 - a > 0$; and $C_J$ becomes vanishingly small at low $\omega$ in the timescale accessible by the MPS time evolution, suggesting a transient-time localization due to translation-invariant flux disorders.
The agreement can be checked explicitly via the continuity equation [Eq.\eqref{eq:appeejj}]. We compute
$\frac{\omega^2}{2-2\cos k}\,C_{\mathcal{E}}(k,\omega)$
using the directly simulated $C_{\mathcal{E}}(k,\omega)$ on the right‐hand side of Eq.\eqref{eq:appeejj}, and compare it to $C_J(k,\omega)$ obtained independently from MPS time evolution.  As shown in Fig.~\ref{fig:ladder2}(b), the two sides coincide to good accuracy, confirming consistency.

\section{Caveats in Self-Consistent Mean-Field Average}
The existence of transient localization as a quasi-DFL in frustrated spin models with fractionalization also highlights a key caveat in applying self‐consistent mean‐field theory to such systems. Because mean‐field self‐averaging, by construction, neglects the nonlocal interference processes that drive Anderson‐type localization, it cannot capture the coherent‐disorder–induced metallic spectrum nor the resulting localization or anomalous transport \cite{Pastor2003,Hadi24}.
This also explains why self-averaging treatments in similar models \cite{ZhangNatComm2022} fail to capture the emergent gapless phase, whereas unbiased numerical methods that include field-induced gauge fluctuations do observe such gapless spectrum \cite{wang2024}.

The key mechanism of the transient-time localization (for suppressed energy conduction) and the gapless Majorana fermions (for large linear specific heat) is the emergent disorder of $Z_2$ gauge fields, leading to the superposition of a random ensemble of flux configurations even at zero temperature. 
It is thus important to point out, the presence of disorder indicates that normal self‐consistent mean‐field theory cannot capture this phenomenon. 
This is because mean‐field self‐averaging, by construction, neglects the nonlocal interference processes that drive Anderson‐type localization.

This can also be made explicitly simple by considering the return probability $R(t)=\bigl|\langle j|e^{-iHt}|j\rangle\bigr|^2$ as an order parameter for localization. 
Here $\ket{j}$ denotes a localized fermion at site $j$, and $R(t)$ measures the memory of that excitation: in a fully localized phase $R(t)$ remains finite at long times, whereas in an extended phase $R(t)$ decays to zero due to destructive interference among many eigenmodes.
Let us start by examining a single disorder realization.  For a localized state $\ket{j}$, the return amplitude is
\begin{equation}
A(t)=\langle j|e^{-iHt}|j\rangle=\sum_{k} |\langle k|j\rangle|^2\,e^{-iE_k t},    
\end{equation}
where $\{|k\rangle\}$ are the eigenstates of $H$ with eigenvalues $E_k$. Because the state is localized, only a few eigenstates have significant weight. This makes dephasing highly unlikely. Hence, for a localized system, $A(t)$ and $R(t) = |A(t)|^2$ do not decay to zero. 

However, upon averaging over flux configurations labeled by $n$ with weights $p_n$, the disorder-averaged return amplitude
\begin{equation}
   \overline{A(t)}=\sum_n p_n\,A_n(t),\;
A_n(t)=\int dE\,\rho_n(E)\,e^{-iEt}
\end{equation}
involves the local density of states (LDOS) $\rho_n(E)=\sum_k|\langle k|j_n\rangle|^2\delta(E-E_k)$.  The ensemble-averaged LDOS $\rho(E)=\sum_n p_n\rho_n(E)$ becomes continuous over the quasiparticle bandwidth D.  Approximating $\rho(E)\approx1/D$ for $|E|\le D/2$ gives
\begin{equation}
\overline{A(t)}\approx\frac{2\sin(\frac{Dt}{2})}{Dt},
\;
R(t)=\bigl|\overline{A(t)}\bigr|^2
=\Bigl[\frac{2\sin(\frac{Dt}{2})}{Dt}\Bigr]^2    
\end{equation}
which decays to zero on a timescale $t\sim2\pi/D$.  Thus, the averaged signal in self-consistent mean-field treatments loses memory on $1/{\rm bandwidth}$ timescales.

A concrete example of this failure in a similar context was recently discussed in Ref.~\cite{Hadi24}, where the authors show that, even when disorder arises classically from thermal boson fluctuations, self‐consistent mean‐field treatments (e.g. dynamical mean-field theory) fail to capture the resulting dynamical localization: by enforcing a local, self‐averaged bath they wash out the spatially correlated fluctuations that give rise to coherent backscattering and thus Anderson‐type localization. In our zero‐temperature setting the disorder instead originates from a superposition of $Z_2$ flux configurations under a moderate magnetic field. Nevertheless, the same caveat applies: mean‐field self‐averaging by construction neglects the nonlocal interference and vertex corrections essential for localization of Majorana fermions, and thus cannot reproduce the suppressed energy transport that we obtain by retaining the full ensemble of gauge backgrounds.

Therefore, capturing both the long‐lived local return amplitude in each sector and the slow decay of the average $R(t)$ requires methods that preserve long-range interference beyond mean‐field self‐averaging, such as those of Ref.~\cite{Pastor2003}; or by unbiased numerical methods such as Lanczos algorithm or MPS dynamics.
This also explain why self-averaging treatments in similar models \cite{ZhangNatComm2022} fail to capture the emergent gapless phase, whereas unbiased numerical methods that include field-induced gauge fluctuations do observe such gapless spectrum \cite{wang2024,penghao24}.

\section{Spreading of Energy Correlation Function}
\begin{figure}[b]
    \centering
    \includegraphics[width=\linewidth]{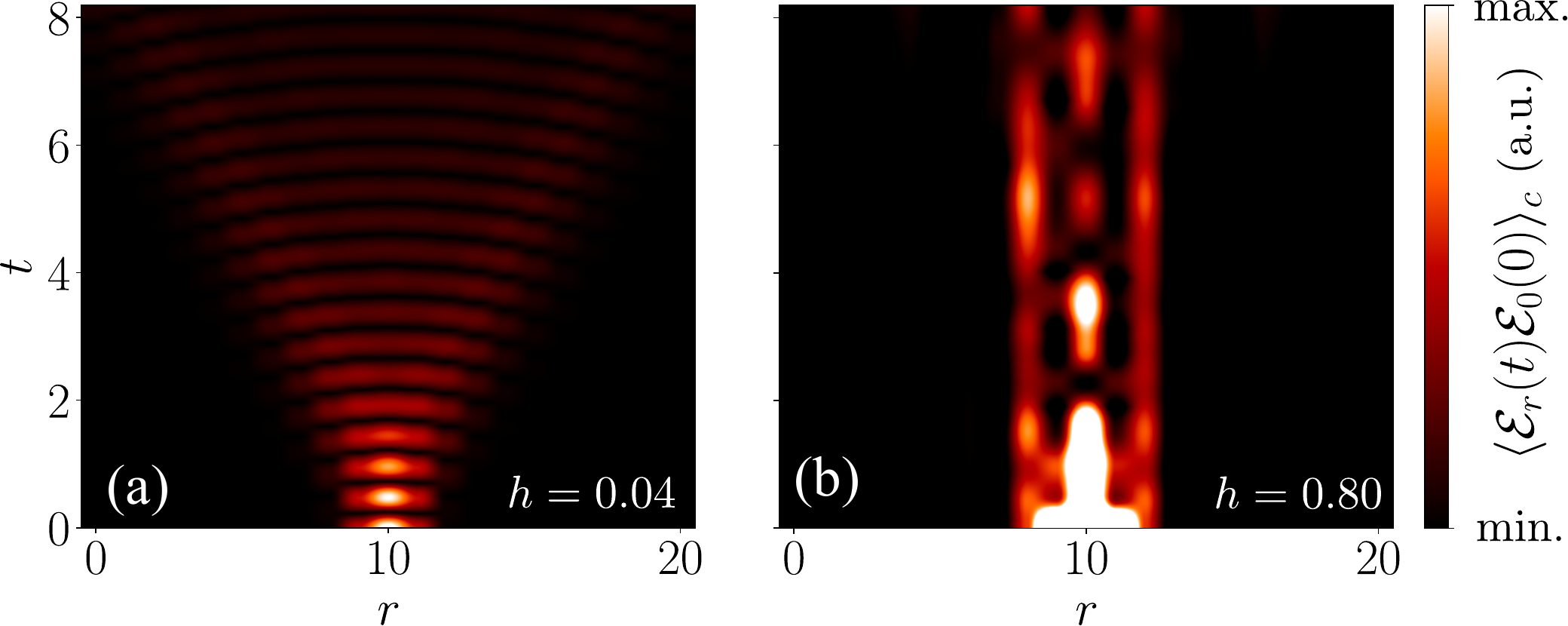}
    \caption{Absolute value of energy correlation function $\expval{0|\mathcal{E}_r(t)\mathcal{E}_{r_0}|0}$ excluding the disconnected ground-state contribution, computed for (a) the spinon band insulator at $h=0.04$ and (b) the gapless and transient localized regime at $h=0.80$.   }
    \label{fig:energycorr}
\end{figure}
In this section we show the energy correlation function in both low and moderate field in supplementary to the quench slow dynamics presented in the main text. We compute the correlation function between local energy densities $C(r, t) = \expval{0|\mathcal{E}_r(t) \mathcal{E}_{r_0}|0}$ by MPS time evolution. 

In Fig.~\ref{fig:energycorr}, we plot the spatio‐temporal energy–density correlation in its absolute value (excluding disconnected ground state contribution)
obtained by MPS time‐evolution for (a) $h=0.04$, deep in the gapped spinon‐insulator phase, and (b) $h=0.80$, in the transiently localized gapless regime.  In (a) the correlation front rapidly expands across the ladder, reflecting propagating gapped excitations, whereas in (b) $C(r,t)$ remains confined to $\lvert r-r_0\rvert\lesssim\xi$ over the entire accessible time window, directly confirming the disorder‐free localization seen in the on‐site energy relaxation discussed in the main text. We emphasize that the entanglement entropy growth during MPS real-time evolution for both $h=0.04$ and $h=0.8$ lies well below the upper bound from the finite bond dimension $\chi=800$, so the confined lightcone in Fig.~\ref{fig:energycorr}(b), as well as slow quench dynamics in Fig.~1(f), are not trivially caused by the capped entropy growth.

The high fidelity of our MPS computation is evidenced by the pronounced rippled structure visible at low field in Fig.~\ref{fig:energycorr}(a).  To understand its origin, we adopt, without loss of generality, a simplified single-Majorana dispersion, $\varepsilon(k)=\cos k + c$,
which captures the essential features of the true band structure, see Fig.~\ref{fig:bands}(a).  Note that the energy density operators defined in Eq.~\eqref{eq:energydensity} are gauge-invariant in the Majorana basis for $h\to 0$, i.e. it can be rewritten as fermion bilinears. Hence, in the low-field limit the two‐point correlator can be written as \cite{wang2024}
\begin{equation}
    C(r,t)\sim\int_{-\pi}^{\pi}\int_{-\pi}^{\pi}dk dq W(k,q)e^{i\bigl[k r - (\varepsilon(k-q)+\varepsilon(q))t\bigr]},
\end{equation}
where $W(k,q)$ is a smooth form factor set by the choice of \(\mathcal{E}_r\).  Assuming $h\rightarrow 0$, $W \to 1$ \cite{wang2024}, and exploiting the trigonometric identity $\varepsilon(k-q)+\varepsilon(q)=2\cos\Bigl(\tfrac k2\Bigr)\cos\Bigl(q-\tfrac k2\Bigr)+2c$, the $q$-integral yields
\begin{equation}
    \int_{-\pi}^{\pi}dq\;e^{-2i t\cos(\tfrac k2)\cos(q-\tfrac k2)}=2\pi\,J_0\bigl(2t\cos\tfrac k2\bigr).
\end{equation}
The remaining $k$-integral then collapses via a second Jacobi–Anger expansion into a single dressed Bessel function of the first kind,
\begin{equation}
    C(r,t)\;\propto\;(-1)^r\,J_{2r}(2t)\,e^{-2ic t},
\end{equation}
so that $|C(r,t)|=2\pi\,\bigl|J_{2r}(2t)\bigr|$. 
Since $J_{2r}(2t)$ oscillates with zeros and maxima that fan out in the $(r,t)$-plane,
one naturally observes the characteristic concentric ripples inside a linear light cone.  These interference fringes directly reflect the coherent superposition of plane‐wave Majorana modes in the integrable limit.
Thus, even in a fully interacting ladder, the retention of an approximate Majorana dispersion near low field ensures that the numerically computed $C(r,t)$ exhibits the same Bessel‐oscillation pattern.  This rippled structure, visible in Fig.~\ref{fig:energycorr}(a), thus serves as a benchmark for our MPS time evolution.

\begin{figure}[t]
    \centering
    \includegraphics[width=\linewidth]{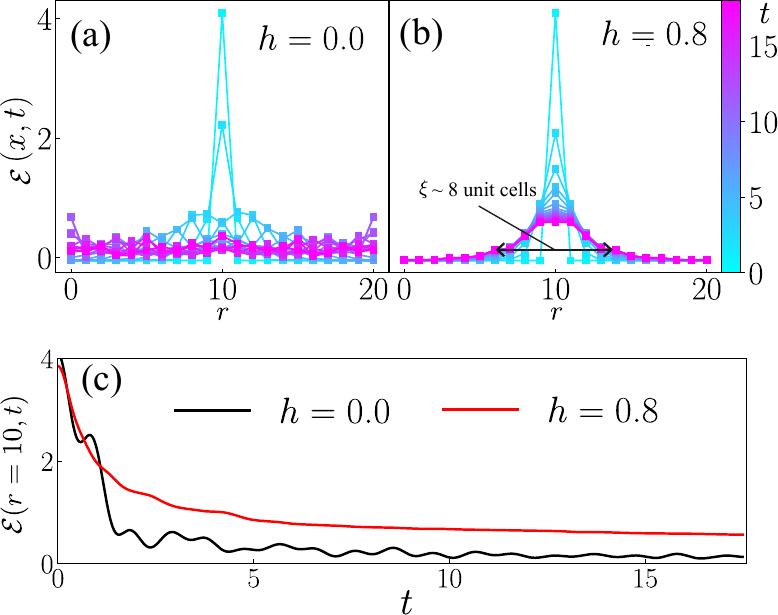}
    \caption{ Cuts of the time evolutions of energy density after a local quench for (a) $h=0.0$ and (b) $h=0.8$ respectively.
    (c) The decay of local energy density $\mathcal{E}(r,t)$ at the perturbed site $r=10$ in a ladder of $L=21$ unit cells (86 sites). Data obtained by MPS simulation with bond dimension $\chi = 800$. }
    \label{fig:profile}
\end{figure}

\section{Additional Results on Quench Dynamics}
\subsection{Evolution of energy density}
In the main text we presented the contour plots of the on‐site energy spreading $\langle \mathcal{E}_i(t)\rangle$ for a quenched state $\ket{0'} = P \ket{0}$, to illustrate slow quench dynamics at low and high fields.  Here, we provide supplementary plots that make the localization length $\xi$ directly visible.

In Fig.~\ref{fig:profile}, we show profile cuts of density spreading,
plotted versus position $r$ at a fixed time slices.  We compare two field values: the gapped spinon‐insulator without localization at $h=0.00$ [Fig.~\ref{fig:profile}(a)] and the transiently localized gapless regime at $h=0.80$ [Fig.~\ref{fig:profile}(b)].
For the gapped spinon insulator in panel (a): The profile remains essentially flat after evolution, confirming that local perturbations extend across the ladder once the front arrives, and at $r=0$ the energy decays rapidly to zero at a time scale $\sim 1/\rm bandwidth$.

For the transient localization regime at $h=0.80$: the energy density profile stopped decaying after settling into a Gaussian-like packet, as clearly shown in Fig.~\ref{fig:profile}(b), where the localization length can be roughly estimated to be $\xi \sim 8$ unit cells. In contrast to the zero-field model, after its initial short-time spreading, the local on-site energy at the perturbed site $r=10$ no longer decays, in the accessible time scale, as shown by the red curve in Fig.~\ref{fig:profile}(c). 
Importantly, this $\xi$ extracted directly from MPS‐evolved data agrees within $O(1)$ precision with the ${\rm IPR} \sim 0.1$ obtained from our effective theory, as shown in Fig.~\ref{fig:ipr}(b) as well as Fig.~3(b) in the main text.
The consistency given by $ \xi \sim 1/\mathrm{IPR}_{\rm eff} \sim 10$
demonstrates that the transient DFL in the gapless regime is quantitatively captured both by numerical simulation and by inverse participation ratio from effective theory.

\subsection{Slow vison dynamics under moderate field}
In the main text we invoke the quasi‐static vison approximation.  This is justified by recent perturbative estimates \cite{Aprem21}, which find the effective vison hopping $t_v$ to be extremely small—of order $\mathcal{O}(10^{-2}h)$ for antiferromagnetic Kitaev exchange due to destructive interference.  For the demonstration field $h\approx0.8$ [before the confinement transition at $h \approx 0.9$], one obtains $t_v\sim10^{-2}$, whereas the Majorana bandwidth remains $O(1)$.  Hence the vison dispersion is nearly flat compared to the fermions, and treating the visons as effectively static is fully valid at low temperatures and over the transient timescales of interest.

Another non-perturbative perspective on the slow visons comes from recent DMRG simulation in Ref.~\cite{Baskaran2023}. Here it was found by DMRG in the Kitaev ladder under a moderate magnetic field that the fidelity $\mathcal{F}(h) = \expval{\psi_0(h)|\psi_0(h+\delta h)}$ goes to zero for $\delta h \ll 1$ at intermediate $h$, signaling a multitude of mutually orthogonal, quasi‐degenerate zero-energy states.  This proliferation of low-energy states has been attributed to a glassy vison sector with slow vison dynamics.

\begin{figure}[t]
    \centering
    \includegraphics[width=0.8\linewidth]{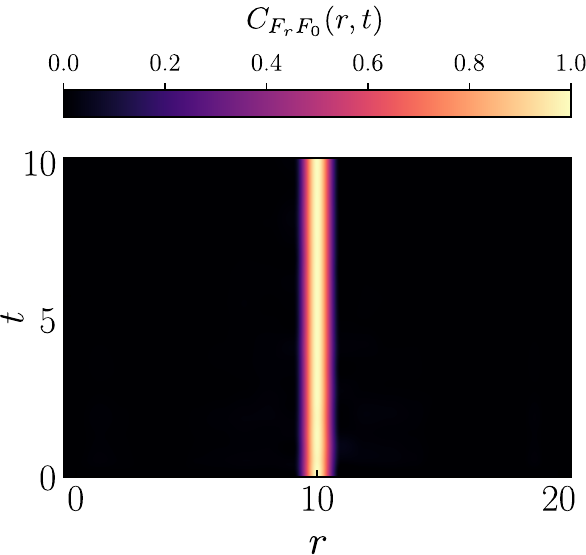}
    \caption{Absolute value of the flux autocorrelation function $C_{F_r F_0}(r, t)$ for $F_a$-flux operator under moderate magnetic field $h=0.80$. }
    \label{fig:flux}
\end{figure}

Here we give further evidence by showing the slow quench dynamics of visons. This is done by computing the flux-flux autocorrelation function after a locally quenched wavefunction $\ket{0'} = P_r \ket{0}$:
\begin{equation} \label{eq:cff}
    C_{F_r F_0}(r, t) = \expval{0'|F_r(t) F_0(t)|0'}_c
\end{equation}
where $F$ is either $F_a$ or $F_b$ flux operator defined in Eq.~\eqref{eq:fluxops}, and $P_r$ a local operator applied at $t=0$. 
We simulate the real-time MPS dynamics in a finite-size ladder up to $100$ site, which is large enough to demonstrate the suppressed spreading of correlation, but small enough so that a finite MPS bond dimension would not limit the accuracy of the variationally obtained ground state. 
Real-time MPS simulation for Eq.~\eqref{eq:cff} is shown in Fig.~\ref{fig:flux}. Remarkably, there is virtually no correlation spreading, further supporting the quasi-static vison ansatz used in the main text besides perturbation theory and speculation from DMRG ground state.  As a result, these slowly fluctuating visons act as intrinsic, self-induced randomness which dynamically localizes the fermions even at zero temperature and near equilibrium.

Furthermore, at lower fields, yet still above the gapped spinon regime, e.g. $h=0.70, \,0.60$, we find that flux dynamics remain as slow over accessible timescales as at $h=0.8$ (Fig.~\ref{fig:flux}).  This suggests that the putative ``flux‐crystal” and ``flux‐glass” phases of Ref.~\cite{Baskaran2023} may in fact be the same phase in the ladders with more than two legs or in 2D thermodynamic limit, with transient localization or sub-diffusion persisting to weaker fields but with an increased localization length due to the reduced flux density.  Indeed, recent infinit-projected-entangled-pair-state (iPEPS) simulations of the 2D Kitaev honeycomb model under a moderate [111] field show that the zero‐energy Majorana modes continuously evolves from a disk around the $\rm M$ point to one around $\rm K$ as the field strength is increased \cite{wang2024,penghao24}.  In a quasi‐1D ladder where momentum resolution is discrete, such a smooth reshaping of the 2D zero‐energy manifold can manifest as successive gap‐closings and gap‐opening transitions, depending on whether a given momentum slice intersects the evolving zero-energy Majorana surface.

\begin{figure*}[t]
    \centering
    \includegraphics[width=0.99\linewidth]{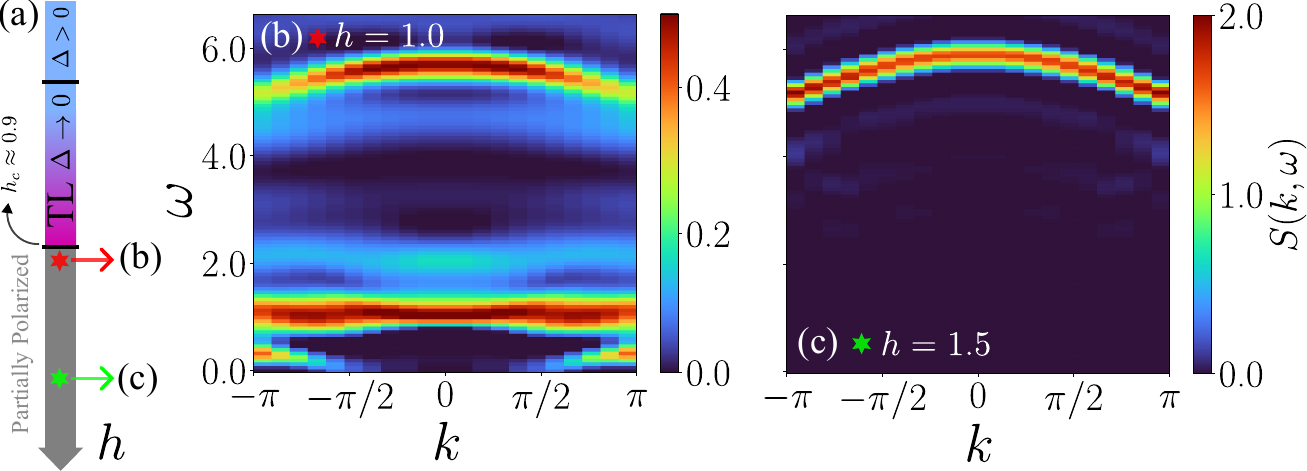}
    \caption{Dynamical spin structure factors $S(k, \omega) = \expval{X_{k}(\omega)X_{-k}(\omega)}$ in the (partially) polarized phase. (a) The schematic phase diagram, marking the computed points in the partially polarized phase in red and green stars. (b) Dynamical spin structure factor at $h=1.0$ just above the polarizing transition $h_c \approx 0.9$. Gapless modes at $k = \pm \pi$ in the TL regime are now gapped out. (c) Dynamical spin structure factor at the high field $h=1.5$.  }
    \label{fig:pp}
\end{figure*}

\section{Additional Results on the Partially Polarized Phase}
Here we show the dynamical spin structure factor at large fields where the system enters the partially polarized phase. 
In Fig.~1 of the main text, we have shown that the system is a gapped spinon band insulator at low field and a gapless phase with transient localization at intermediate fields before the polarizing transition $h_c \approx 0.9$. Beyond $h_c$ the fractionalized excitations become confined and the spins partially polarized with gapped excitations.  
As is shown in Fig.~\ref{fig:pp}, at $h=1.0$ immediately above the polarizing transition $h_c \approx 0.9$, the otherwise gapless points at $k = \pm \pi$ in the intermediate phase are gapped out in the partially polarized phase [cf. Fig.~1(d) in the main text], and the modes become sharper than that in the intermediate phase, albeit with some continuum signals at higher energies $\omega \in (0.5, 2.5)$, as shown in Fig.~\ref{fig:pp}(b), due to its proximity to the critical point. The gapped continuum at low energies becomes even more gapped and eventually disappears at higher magnetic field, where the system is to be described by a spin wave theory with sharp magnon modes, as shown in Fig.~\ref{fig:pp}(c).

\begin{figure}[b]
    \centering
    \includegraphics[width=\linewidth]{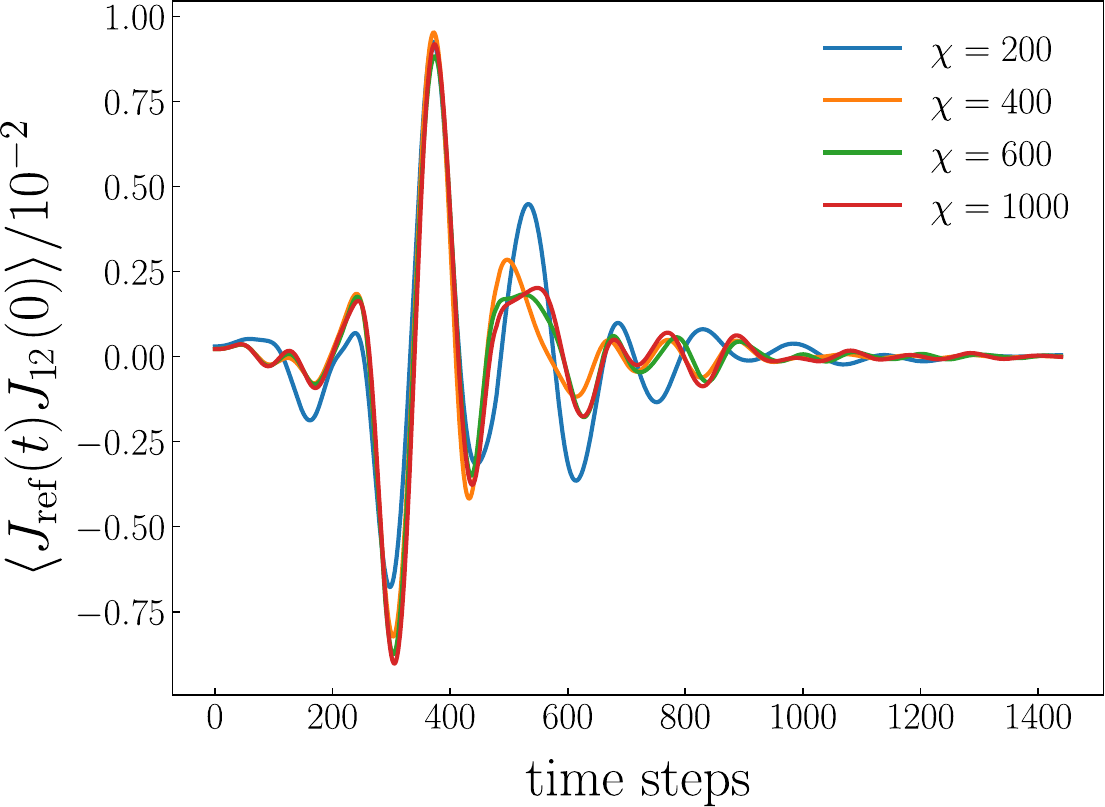}
    \caption{The real-time-resolved current-current correlation between $J$ at a site of reference ($J_{\rm ref}$) and $J_{12}$ at the $12$-th unit cell. Time step is set to $\delta t = 0.01$ measured in the unit of exchange interaction strength.    }
    \label{fig:converge}
\end{figure}

\section{Convergence of MPS Simulations}
In this section, we demonstrate the numerical convergence of our MPS simulations.  We obtained ground states by two-site DMRG on open-boundary ladders up to 25‐unit‐cell (100‐site) using a bond dimension of $\chi=1000$.  
The energy converged at the order of $10^{-5}$, and the truncation error is bounded by $10^{-10}$ at moderate field and is even smaller at low field. 
The most computationally demanding step is the real‐time evolution, which we perform using the W-II MPO approximation scheme of Ref.~\cite{Zaletel15}, as implemented in the TeNPy library \cite{tenpy2024} for quench dynamics or conductivity, and the Krylov vector-space targeting approach to correlation vector with the same bond dimension for dynamical spin structure factors \cite{White99,Alvarez16}.  To benchmark the convergence in dynamical spectra, we calculate 
the time-resolved current–current correlation shown in Fig.~\ref{fig:converge}, increasing $\chi$ from 200 to 400 improves the accuracy, but beyond this point the changes are minor and convergence is essentially achieved. For example, raising $\chi$ further from 600 to 1000 yields only a negligible gain.
Based on this, real-time evolution presented in the main text was obtained on an $86$-site ladder with $\chi=800$ and $100$-site ladder with $\chi=1000$, which lies well within the converged regime while optimizing computational cost.

\end{document}